\documentclass[11pt]{article}

\usepackage{epsf}
\usepackage{psfig}
\usepackage[dvips]{epsfig}
\usepackage{amsmath}
\usepackage{amssymb}
\usepackage{fullpage}

 \advance \textwidth by -2in
 \advance \textheight by -3in
\def\le{\left(}
\def\ri{\right)}
\def\les{\left[}
\def\ris{\right]}

\def\ric{\right\}}

\def\c#1{\cite{#1}}

\def\r#1{(\ref{#1})}

\def\epso{\epsilon_o}

\def\muo{\mu_o}
\def\lambdao{\lambda_o}
\def\etao{\eta_o}
\def\ko{k_o}

\def\sp{\bf s}
\def\pinc{{\bf p}_+}
\def\pref{{\bf p}_-}

\def\einc{{\bf e}_{inc}(z)}
\def\erefl{{\bf e}_{ref}(z)}
\def\etr{{\bf e}_{tr}(z)}
\def\hinc{{\bf h}_{inc}(z)}
\def\hrefl{{\bf h}_{ref}(z)}
\def\htr{{\bf h}_{tr}(z)}

\def\cphi{\cos\phi}
\def\sphi{\sin\phi}
\def\ctheta{\cos\theta}

\def\aL{a_L}
\def\aR{a_R}
\def\rL{r_L}
\def\rR{r_R}
\def\tL{t_L}
\def\tR{t_R}

\def\rRR{r_{RR}}
\def\rRL{r_{RL}}
\def\rLR{r_{LR}}
\def\rLL{r_{LL}}

\def\tRR{t_{RR}}
\def\tRL{t_{RL}}
\def\tLR{t_{LR}}
\def\tLL{t_{LL}}

\def\RRR{R_{RR}}
\def\RRL{R_{RL}}
\def\RLR{R_{LR}}
\def\RLL{R_{LL}}

\def\TRR{T_{RR}}
\def\TRL{T_{RL}}
\def\TLR{T_{LR}}
\def\TLL{T_{LL}}

\def\Ezdc{E_z^{dc}}

\def\dprime{{\prime\prime}}

\def\ux{\hat{\bf{u}}_x}
\def\uy{\hat{\bf{u}}_y}
\def\uz{\hat{\bf{u}}_z}

\begin{document}

\begin{center}
{\large {\bf Theory of electrically controlled exhibition of circular
Bragg phenomenon by an obliquely excited structurally chiral material}}
\vskip 0.5cm

\noindent  { Akhlesh Lakhtakia}\footnote{E--mail: akhlesh@psu,edu}
\vskip 0.2cm
\noindent {\em Computational \& Theoretical Materials Sciences Group (CATMAS)\\
Department of Engineering Science \& Mechanics\\
Pennsylvania State University, University Park, PA 16802--6812, USA}
\vskip 0.5cm

\noindent{Juan Adrian Reyes}\footnote{E--mail: adrian@fisica.unam.mx}
\vskip 0.2cm
{\em Instituto de Fisica\\ 
Universidad Nacional Autonoma de Mexico\\ 
Apartado Postal 20--364, C.P. 01000, Mexico D.F., Mexico}

\end{center}

\vskip 0.5cm

\noindent {\bf Abstract:}
The boundary--value problem of the reflection and transmission of a
plane wave due to a
slab  of an electro--optic structurally chiral material (SCM) is formulated in terms
of a  4$\times$4 matrix ordinary
differential equation.
The SCM  slab can be locally endowed with one of 20 classes
of
point group symmetry, and is subjected to a dc voltage across its thickness. The enhancement (and, in some cases, the production) of the circular Bragg phenomenon (CBP) by the application of the dc voltage has either switching or 
circular--polarization--rejection
applications in optics. The twin possibilities of thinner filters and
electrical manipulation
of the CBP, depending on the local crystallographic class as well as 
the constitutive parameters of the SCM, emerge.


\section{Introduction}

The Bragg phenomenon is exhibited by a slab of a material whose electromagnetic
constitutive properties are  periodically nonhomogeneous in the thickness direction. Its signature
is very high reflectance  in a certain wavelength--regime, provided
the slab is thick enough to have a sufficiently large  number of
periods. This phenomenon is commonly exploited to make dielectric
mirrors  in optics \cite{Macl,Baum}. 

If the material is isotropic, no dependence
of the Bragg phenomenon
on the polarization state of a normally incident electromagnetic wave is evident.
The material must be anisotropic for the Bragg phenomenon to
discriminate between two mutually orthogonal polarization states \cite{HW-book}.

Periodicity arises from structural chirality~---~i.e., a heliocidal variation
of anisotropy along a fixed axis~---~in
cholesteric liquid crystals \cite{DeGennes,APRO} and chiral sculptured thin films
\cite{LakhtakiaB,RBL}, which exemplify
structurally chiral materials (SCMs). Both types of SCMs are continuously
nonhomogeneous in the thickness direction. SCMs can also
be piecewise continuous, as proposed about 140 years ago by Reusch \cite{Reusch}
and expanded upon recently by Hodgkinson {\em et al.} \cite{HLWDM}.
As the periodicity arises from structural chirality, incident electromagnetic plane
waves of left--
and right--circular polarization (LCP and RCP) states are reflected and transmitted
differently in the Bragg wavelength--regime, and the Bragg phenomenon
is then called the {\it circular} Bragg phenomenon (CBP). Exhibition of the
CBP by cholesteric liquid crystals and  chiral sculptured thin films
underlies their use as circular--polarization rejection filters in optics \cite{DeGennes,LakhtakiaB,Jacobs}.

Control of the CBP is very desirable for tuning the Bragg regime
as well as for switching applications. One way would be to use SCMs that are
electro--optic, because then the CBP could be electrically controlled. This
possibility, also suggested by the
fabrication of electro--optic Sol\u{c} filters \cite{Pin}, was proposed
and theoretically examined by us in a recent publication \cite{RL06}.
Therein, the SCM was assumed
to possess locally a $\bar{4}2m$ point group symmetry for the
exhibition of the Pockels effect \cite{Boyd}, and the electromagnetic wave
was taken to be normally incident on the SCM slab across whose thickness
a low--frequency (or dc) electric field was supposed to be applied. The Pockels effect
was found to enhance the CBP \cite{RL06,RLno2}, so much so that
it could
engender the CBP even if that phenomenon were to absent in the absence of a 
dc electric field. 

In this paper, we take a comprehensive look at the planewave response
characteristics of an electro--optic SCM slab. The electromagnetic plane wave can be either
normally or obliquely incident. The SCM slab is locally endowed with one of 20 classes
of
point group symmetry \cite{Boyd} relevant to the excitation
of the Pockels effect by a dc voltage applied across its thickness.

The organization of this paper is as follows:  The theoretical formulation is presented
in Section \ref{the}, beginning with the optical relative permittivity matrixes of
a homogeneous electro--optic material and a SCM, going on to
exploit the Oseen transformation to derive a 4$\times$4 matrix ordinary
differential equation for electromagnetic propagation in a SCM, then setting up a boundary--value
problem to compute the reflectances and transmittances of a SCM slab,
and finally discussing the salient features of axial propagation in a SCM. Section
\ref{nrd} is devoted to the presentation and discussion of numerical results.
 CBP enhancement   by the application of the dc voltage is 
 shown to have either switching or circular--polarization--rejection
applications. The possibilities of thinner filters and electrical manipulation
of the CBP, depending on the local crystallographic class as well as 
the constitutive parameters of the SCM, emerge from analysis for
normal incidence.

A note about notation: Vectors are denoted in boldface; the cartesian unit vectors are represented
by $\hat{\mathbf{u}}_x$, $\hat{\mathbf{u}}_y$, and $\hat{\mathbf{u}}_z$;
symbols for column vectors and matrixes are decorated by an overbar; and an $%
\exp(-i\omega t)$ time--dependence is implicit with $\omega$ as the angular
frequency.

\section{Theoretical formulation \label{the}}

We are interested in the reflection and transmission of plane waves due to
a SCM slab of thickness $L$. The axis of structural chirality
 of the SCM
is designated as the $z$ axis, and the SCM is subjected to a dc
electric field $\mathbf{E}^{dc} = E_z^{dc}\,\hat{\mathbf{u}}_z$.
The half--spaces $z\leq 0$ and
$z\geq L$ are vacuous.
An arbitrarily polarized plane wave 
is obliquely incident 
on the SCM from the half--space
$z\leq 0$. As a result, reflected and transmitted plane waves exist
in the half--spaces $z\leq 0$ and
$z\geq L$, respectively. A boundary--value problem has to be solved
in order to determine the reflection and transmission coefficients.

\subsection{Pockels Effect}
In order to delineate the electro--optic properties of the chosen SCM,
let us first consider a (nondissipative) homogeneous dielectric material susceptible to the
Pockels effect when subjected to a dc field $\mathbf{E}^{dc}$.
The reciprocal of the optical relative permittivity matrix is usually
reported in the literature as \cite{Boyd} 
\begin{equation}
\displaystyle{\ \bar{\epsilon}_{PE}^{-1}=\left( 
\begin{array}{ccc}
1/\epsilon _{1}^{(0)}+\sum_{K=1}^3 r_{1K}E_{K}^{dc} & \,\sum_{K=1}^3
r_{6K}E_{K}^{dc} & \,\sum_{K=1}^3 r_{5K}E_{K}^{dc} \\[5pt] 
\sum_{K=1}^3 r_{6K}E_{K}^{dc} & \,1/\epsilon _{2}^{(0)}+\sum_{K=1}^3
r_{2K}E_{K}^{dc} & \,\sum_{K=1}^3 r_{4K}E_{K}^{dc} \\[5pt] 
\sum_{K=1}^3 r_{5K}E_{K}^{dc} & \,\sum_{K=1}^3 r_{4K}E_{K}^{dc} & 
\,1/\epsilon _{3}^{(0)}+\sum_{K=1}^3 r_{3K}E_{K}^{dc}
\end{array}
\right) }  \label{Poc}
\end{equation}
in the principal Cartesian coordinate system (with axes labeled 1, 2, and 3)
relevant to the crystallographic structure of the material \cite[Table 7.1]
{Auld}. Here, $E_{1,2,3}^{dc}$ are the Cartesian components of the dc
electric field, $\epsilon _{1,2,3}^{(0)}$ are the principal relative
permittivity scalars in the optical regime, whereas $r_{JK}$ (with $1\leq
J\leq 6$ and $1\leq K\leq 3$) are the electro--optic coefficients in the
traditional contracted or abbreviated notation for representing symmetric
second--order tensors \cite{Boyd,Auld}. 

This material can be isotropic, uniaxial, or biaxial, depending on the relative
values of
$\epsilon_1^{(0)}$, $\epsilon_2^{(0)}$, and $\epsilon_3^{(0)}$. Furthermore,
this material may belong to one of  20 crystallographic classes of point group
symmetry, in accordance with the relative
values of the electro--optic coefficients $r_{JK}$ \cite[pp. 46--47]{Boyd}.

Correct to the first order in the components of the dc electric field, we
get the linear approximation 
\begin{equation}
\displaystyle{\ \bar{\epsilon}_{PE}\approx\left( 
\begin{array}{ccc}
\epsilon _{1}^{(0)}(1-\epsilon _{1}^{(0)}\sum_{K=1}^3 r_{1K}E_{K}^{dc} ) & 
-\epsilon _{1}^{(0)}\epsilon _{2}^{(0)}\sum_{K=1}^3 r_{6K}E_{K}^{dc} & 
-\epsilon _{1}^{(0)}\epsilon _{3}^{(0)}\sum_{K=1}^3 r_{5K}E_{K}^{dc} \\[5pt] 
-\epsilon _{2}^{(0)}\epsilon _{1}^{(0)}\sum_{K=1}^3 r_{6K}E_{K}^{dc} & 
\epsilon _{2}^{(0)}(1-\epsilon _{2}^{(0)}\sum_{K=1}^3 r_{2K}E_{K}^{dc} ) & 
-\epsilon _{2}^{(0)}\epsilon _{3}^{(0)}\sum_{K=1}^3 r_{4K}E_{K}^{dc} \\[5pt] 
-\epsilon _{3}^{(0)}\epsilon _{1}^{(0)}\sum_{K=1}^3 r_{5K}E_{K}^{dc} & 
-\epsilon _{3}^{(0)}\epsilon _{2}^{(0)}\sum_{K=1}^3 r_{4K}E_{K}^{dc} & 
\epsilon _{3}^{(0)}(1-\epsilon _{3}^{(0)}\sum_{K=1}^3 r_{3K}E_{K}^{dc} )
\end{array}
\right) }  \label{PocEps}
\end{equation}
from (\ref{Poc}).

\subsection{Structurally chiral material}

As the electro--optic SCM has the $z$ axis as its axis of
chiral nonhomogeneity and is subjected to a dc
electric field $\mathbf{E}^{dc} = E_z^{dc}\,\hat{\mathbf{u}}_z$,
the optical relative permittivity matrix of this material may be stated as 
\begin{equation}
\bar{\epsilon}^{SCM}(z) = \bar{S}_{z}\left(\frac{h\pi z}{\Omega}\right)\cdot%
\bar{R}_{y}(\chi) \cdot\bar{\epsilon}_{PE} \cdot\bar{R}_{y}(\chi)\cdot \bar{S%
}_{z}\left(-\,\frac{h\pi z}{\Omega}\right)\,,  \label{AAepsr}
\end{equation}
where $\bar{\epsilon}_{PE}$ is specified by (\ref{PocEps}). The tilt matrix 
\begin{equation}
\bar{R}_{y}(\chi )=\left( 
\begin{array}{ccc}
-\sin \chi & 0 & \cos \chi \\ 
0 & -1 & 0 \\ 
\cos \chi & 0 & \sin \chi
\end{array}
\right)
\end{equation}
involves the angle $\chi \in\left[0,\pi/2\right]$ with respect to the $x$
axis in the $xz$ plane. The use of the rotation matrix 
\begin{equation}
\bar{S}_z(\zeta)=\left( 
\begin{array}{ccc}
\cos \zeta & -\,\sin\zeta & 0 \\ 
\sin\zeta & \cos \zeta & 0 \\ 
0 & 0 & 1
\end{array}
\right)
\end{equation}
in (\ref{AAepsr}) involves the half--pitch $\Omega $ of the SCM along the $z$
axis. In addition, the handedness parameter $h=1$ for structural
right--handedness and $h=-1$ for structural left--handedness. Depending on
the relationships between $\epsilon_1^{(0)}$, $\epsilon_2^{(0)}$, and $\epsilon_3^{(0)}$,
a SCM may be classified as locally isotropic, locally uniaxial, or locally biaxial~---~the
qualifier {\it local} referring to the crystallographic symmetry in any plane $z=\mbox{constant}$.

Furthermore, for the specific configuration of the dc electric field, we get 
\begin{equation}
\left. 
\begin{array}{l}
E_{1}^{dc}= E_z^{dc}\,\cos \chi \\[5pt] 
E_{2}^{dc}=0 \\[5pt] 
E_{3}^{dc}= E_z^{dc}\,\sin \chi
\end{array}
\right\} \,.
\end{equation}

\subsection{Propagation in a SCM}

The Maxwell curl postulates for the chosen SCM are given by 
\begin{eqnarray}
\nonumber
&&\left. 
\begin{array}{l}
\nabla \times \mathbf{E}(x,y,z)=i\omega\mu_o\mathbf{H}(x,y,z) \\[5pt] 
\nabla \times \mathbf{H}(x,y,z)=-i\omega\epsilon_o\bar{\epsilon}%
^{SCM}(z)\cdot \mathbf{E}(x,y,z)
\end{array}
\right\} \,,
\\
&&\qquad\qquad 0<z<L\,,
\end{eqnarray}
where $\epsilon_o$ and $\mu_o$ are the permittivity and the permeability of
free space (i.e., vacuum).

As a plane wave is  incident obliquely on the SCM,   $\forall z$ we set \cite{LW97,VLprsa} 
\begin{equation}
\left. 
\begin{array}{l}
\mathbf{E}(x,y,z)= \mathbf{e}(z)\, \exp\left[
i\kappa(x\cos\phi+y\sin\phi)\right] \\[5pt] 
\mathbf{H}(x,y,z)= \mathbf{h}(z)\, \exp\left[
i\kappa(x\cos\phi+y\sin\phi)\right]
\end{array}
\right\}\,,
\end{equation}
where the wavenumber $\kappa$ and the angle $\phi$ are determined by the
incidence conditions. The essential part of the Maxwell curl postulates can
then be stated in terms of the column vector \cite{LW97,Marcuvitz} 
\begin{equation}
{\bar{\psi}}\left( z\right) =\left( 
\begin{array}{c}
e_{x}(z) \\ 
e_{y}(z) \\ 
h_{x}(z) \\ 
h_{y}(z)
\end{array}
\right) \, .  \label{campoe_h}
\end{equation}

Inside the SCM, it is advantageous to exploit the Oseen transformation \cite
{LW97,Oseen} by defining the column vector 
\begin{equation}
{\bar{\psi}}^{\prime }(z)=\bar{M}\left(\frac{h\pi z}{\Omega}\right)\cdot {%
\bar{\psi}}(z)\,,
\end{equation}
where the unitary 4$\times $4 matrix 
\begin{equation}
\bar{M}(\zeta)=\left( 
\begin{array}{cccc}
\cos \zeta & \sin \zeta & 0 & 0 \\ 
-\sin \zeta & \cos \zeta & 0 & 0 \\ 
0 & 0 & \cos \zeta & \sin \zeta \\ 
0 & 0 & -\sin \zeta & \cos \zeta
\end{array}
\right) \,.
\end{equation}
Following the procedure outlined by Lakhtakia and Weiglhofer \cite{LW97}, we
have established that ${\bar{\psi}}^{\prime }(z)$ satisfies the matrix
ordinary differential equation 

\begin{equation}  \label{oblique}
\frac{d}{dz}{\bar{\psi}}^{\prime }(z)= i \bar{A}^\prime(z)\cdot{\bar{\psi}}%
^{\prime }(z)\,, \qquad 0 < z <L\,,
\end{equation}
where 
\begin{eqnarray}
\bar{A}^\prime(z) &=& \left( 
\begin{array}{cccc}
0 & -i\frac{h\pi}{\Omega} & 0 & \omega\mu_o \\ 
i\frac{h\pi}{\Omega} & 0 & -\omega\mu_o & 0 \\ 
0 & -\omega\epsilon_o\epsilon_2^{(0)} & 0 & -i\frac{h\pi}{\Omega} \\ 
\omega\epsilon_o\epsilon_d & 0 & i\frac{h\pi}{\Omega} & 0
\end{array}
\right)  \nonumber 
+\kappa\delta_\epsilon \left( 
\begin{array}{cccc}
\cos u & 0 & 0 & 0 \\ 
-\sin u & 0 & 0 & 0 \\ 
0 & 0 & 0 & 0 \\ 
0 & 0 & \sin u & \cos u
\end{array}
\right)  \nonumber \\[6pt]
&+& \frac{\kappa^2}{\omega\epsilon_o}\,\frac{\epsilon_d}{\epsilon_1^{(0)}%
\epsilon_3^{(0)}} \left(1+\frac{\alpha_2}{\epsilon_1^{(0)}\epsilon_3^{(0)}}%
\right) \left( 
\begin{array}{cccc}
0 & 0 & -\sin u \cos u & -\cos^2u \\ 
0 & 0 & \sin^2u & \sin u \cos u \\ 
0 & 0 & 0 & 0 \\ 
0 & 0 & 0 & 0
\end{array}
\right)  \nonumber \\[6pt]
&+& \frac{\kappa^2}{\omega\mu_o} \left( 
\begin{array}{cccc}
0 & 0 & 0 & 0 \\ 
0 & 0 & 0 & 0 \\ 
\sin u \cos u & \cos^2u & 0 & 0 \\ 
- \sin^2u & - \sin u \cos u & 0 & 0
\end{array}
\right)  \nonumber \\[6pt]
&-&\omega\epsilon_o\frac{\epsilon_2^{(0)}}{\epsilon_1^{(0)}} \left( 
\begin{array}{cccc}
0 & 0 & 0 & 0 \\ 
0 & 0 & 0 & 0 \\ 
\epsilon_e+\epsilon_h & -\epsilon_m & 0 & 0 \\ 
\epsilon_\iota\cos\chi+(\epsilon_j+\epsilon_\ell) \frac{\sin 2\chi}{2}%
+\epsilon_k\sin\chi & -(\epsilon_e+\epsilon_h) & 0 & 0
\end{array}
\right)  \nonumber \\[6pt]
&+&\kappa\frac{\epsilon_2^{(0)}}{\epsilon_1^{(0)}\epsilon_3^{(0)}} \left( 
\begin{array}{cccc}
-\frac{\alpha_1\cos u}{\epsilon_1^{(0)}} & - {(\epsilon_f+\epsilon_g)\cos u}
& 0 & 0 \\ 
\frac{\alpha_1\sin u}{\epsilon_1^{(0)}} & {(\epsilon_f+\epsilon_g)\sin u} & 0
& 0 \\ 
0 & 0 & {(\epsilon_f+\epsilon_g)\sin u} & {(\epsilon_f+\epsilon_g)\cos u} \\ 
0 & 0 & -\frac{\alpha_1\sin u}{\epsilon_1^{(0)}} & -\frac{\alpha_1\cos u}{%
\epsilon_1^{(0)}}
\end{array}
\right)\,,
\end{eqnarray}

\begin{eqnarray}  
\nonumber
&&\alpha_1= \epsilon _{1}^{(0)}\epsilon_j\cos^2\chi-\epsilon
_{3}^{(0)}\epsilon_\ell\sin^2\chi +\epsilon _{1}^{(0)}\epsilon_k\cos\chi\\
&&\qquad\quad
-\epsilon _{3}^{(0)}\epsilon_\iota\sin\chi\,, \\
\nonumber
&&\alpha_2=\left(\epsilon _{1}^{(0)}\epsilon_n+\epsilon
_{3}^{(0)}\epsilon_p\right)\cos\chi \\
&&\qquad\quad+\left(\epsilon
_{1}^{(0)}\epsilon_s+\epsilon _{3}^{(0)}\epsilon_q\right)\sin\chi\,, \\[6pt]
&&\delta_\epsilon =\epsilon _{d}\sin 2\chi \frac{\left( \epsilon
_{1}^{(0)}-\epsilon _{3}^{(0)}\right) }{2\epsilon _{1}^{(0)}\epsilon
_{3}^{(0)}} \\[9pt]
\label{epsd-def}
&&\epsilon _{d}=\frac{\epsilon _{1}^{(0)}\epsilon _{3}^{(0)}}{\epsilon
_{1}^{(0)}\cos ^{2}\chi +\epsilon _{3}^{(0)}\sin ^{2}\chi }\,, \\[9pt]
&&\epsilon_{e} = E_z^{dc} \epsilon_1^{(0)} \epsilon_d
(r_{41}\cos^2\chi-r_{63}\sin^2\chi)\,, \\[9pt]
&&\epsilon_{f}=E_z^{dc}\epsilon_d\sin\chi\,\cos\chi(r_{41}%
\epsilon_3^{(0)}+r_{63}\epsilon_1^{(0)})\,, \\
&&\epsilon_{g} = E_z^{dc} \epsilon_d
(r_{43}\epsilon_3^{(0)}\sin^2\chi+r_{61}\epsilon_1^{(0)}\cos^2\chi)\,, \\
&&\epsilon_{h}=E_z^{dc}
\epsilon_1^{(0)}\epsilon_d\sin\chi\,\cos\chi(r_{43}-r_{61})\,, \\
&&\epsilon_\iota=E_z^{dc} \frac{\epsilon_1^{(0)}}{\epsilon_2^{(0)}}%
\,\epsilon_d^2 (r_{31}\cos^2\chi-r_{53}\sin^2\chi)\,, \\
&&\epsilon_{j}=E_z^{dc} \frac{\epsilon_1^{(0)}}{\epsilon_2^{(0)}}%
\,\epsilon_d^2 \sin\chi (r_{11} -r_{53} )\,, \\
&&\epsilon_{k}=E_z^{dc} \frac{\epsilon_1^{(0)}}{\epsilon_2^{(0)}}%
\,\epsilon_d^2 (r_{13}\sin^2\chi-r_{51}\cos^2\chi)\,, \\
&&\epsilon_{\ell}=E_z^{dc} \frac{\epsilon_1^{(0)}}{\epsilon_2^{(0)}}%
\,\epsilon_d^2 \cos\chi (r_{33} -r_{51} )\,, \\
&&\epsilon_{m}=E_z^{dc}\epsilon_1^{(0)}\epsilon_2^{(0)}
(r_{21}\cos\chi+r_{23}\sin\chi)\,, \\
&&\epsilon_{n} = E_z^{dc} \epsilon_d
(r_{53}\epsilon_3^{(0)}\sin^2\chi+r_{11}\epsilon_1^{(0)}\cos^2\chi)\,, \\
&&\epsilon_{p}=E_z^{dc}\epsilon_d\sin^2\chi\,
(r_{31}\epsilon_3^{(0)}+r_{53}\epsilon_1^{(0)})\,, \\
&&\epsilon_{q} = E_z^{dc} \epsilon_d
(r_{33}\epsilon_3^{(0)}\sin^2\chi+r_{51}\epsilon_1^{(0)}\cos^2\chi)\,, \\
&&\epsilon_{s}=E_z^{dc}\epsilon_d\cos^2\chi\,
(r_{51}\epsilon_3^{(0)}+r_{13}\epsilon_1^{(0)})\,, \\ 
&&u = \frac{h\pi z}{\Omega}\,-\phi\,.
\end{eqnarray}
The matrix $\bar{A}^\prime(z)$ is stated correct to the first order in $\Ezdc$.

By virtue of linearity, the solution of the 4$\times$4 matrix ordinary differential
equation \r{oblique}
must be of the form
\begin{equation}  \label{oblique-soln1}
{\bar{\psi}}^{\prime }(z_2)= \bar{U}^\prime(z_2-z_1)\cdot{\bar{\psi}}%
^{\prime }(z_1)\,,   
\end{equation}
whence
\begin{eqnarray}
\nonumber
\bar{\psi}(z_2)&=&\bar{M}\left(-\,\frac{h\pi z_2}{\Omega}\right)\cdot\bar{U}^\prime(z_2-z_1)\cdot\bar{M}\left(\frac{h\pi z_1}{\Omega}\right)\cdot\bar{\psi}%
(z_1)\\[4pt]
\nonumber
&\equiv&\bar{U}(z_2-z_1)\cdot\bar{\psi}(z_1)\,,\\
&&\qquad \quad0\leq z_\ell\leq L\,,\quad
\ell=1,2\,. \label{oblique-soln2}
\end{eqnarray}
There are at least two methods for calculating $ \bar{U}^\prime(z)$ \cite{LW97,SchHer,Polo2},
and we chose to implement the piecewise homogeneity approximation
method \cite{LakhtakiaB,Polo2}.

\subsection{Reflection and transmission by a SCM slab}

The incident plane wave is  delineated by the phasors \cite{LakhtakiaB,VLprsa}
\begin{eqnarray}
&&
\nonumber
\left.\begin{array}{l}
\einc= 
\le \aL\,\frac{i\sp-\pinc}{\sqrt{2}} -
\aR\,\frac{i\sp+\pinc}{\sqrt{2}} \ri\,e^{i\ko z\cos\theta}
\\[14pt]
\hinc= 
-i\etao^{-1}\,\le \aL\,\frac{i\sp-\pinc}{\sqrt{2}} +
\aR\,\frac{i\sp+\pinc}{\sqrt{2}} \ri\,e^{i\ko z\cos\theta}
\end{array}\ric
\,,
\\
&&
\qquad\quad z \leq 0\,,
\label{eq9.50}
\end{eqnarray}
where 
$\etao=\sqrt{\muo/\epso}$ is the intrinsic
impedance of free space;
 $\aL$ and $\aR$ are the
amplitudes of the  LCP
and RCP components, respectively; and
the vectors
\begin{eqnarray}
\label{eq9.51}
&&\sp=-\ux\sin\phi + \uy \cos\phi\,,
\\
\label{eq9.52}
&&{\bf p}_\pm=\mp\le \ux \cos\phi + \uy \sin\phi \ri \cos\theta + 
\uz \sin\theta\,
\end{eqnarray}
are of unit magnitude. The propagation vector of the
 incident  
plane wave 
makes
an angle $\theta \in \les 0,\,\pi/2\ri$ with respect to the $+z$ axis,
and is inclined to the $x$ axis in the $xy$ plane
by an angle $\psi \in\les 0,\,2\pi\ris$; accordingly, the transverse
wavenumber 
\begin{equation}
\kappa = \ko\,\sin\theta\,,
\end{equation}
where $\ko=\omega\sqrt{\epso\muo}$ is the wavenumber in free space.
The free--space wavelength is denoted by $\lambdao=2\pi/\ko$.

The electromagnetic field phasors associated with
the reflected and 
transmitted plane waves, respectively,  
are expressed by \cite{LakhtakiaB,VLprsa}
\begin{eqnarray}
\nonumber
&&\left.\begin{array}{l}
\erefl= 
\le -\rL\,\frac{i\sp-\pref}{\sqrt{2}} +
\rR\,\frac{i\sp+\pref}{\sqrt{2}} \ri\,e^{-i\ko z\cos\theta}
\\[14pt]
\hrefl= 
i\etao^{-1}\,\le \rL\,\frac{i\sp-\pref}{\sqrt{2}} +
\rR\,\frac{i\sp+\pref}{\sqrt{2}} \ri\,e^{-i\ko z\cos\theta}
\end{array}\ric
\,,\\
&&\qquad
\quad z \leq 0\,,
\label{eq9.53}
\end{eqnarray}
and
\begin{eqnarray}
\nonumber
&&
\left.\begin{array}{l}
\etr= 
\le \tL\,\frac{i\sp-\pinc}{\sqrt{2}} -
\tR\,\frac{i\sp+\pinc}{\sqrt{2}} \ri\,\\
\qquad\qquad
\times
e^{i\ko (z-L)\cos\theta}
\\[14pt]
\htr= 
-i\etao^{-1}\,\le \tL\,\frac{i\sp-\pinc}{\sqrt{2}} +
\tR\,\frac{i\sp+\pinc}{\sqrt{2}} \ri\,\\ 
\qquad\qquad 
\times 
e^{i\ko (z-L)\cos\theta}
\end{array}\ric
\,,\\
&&\qquad
\quad z \geq L\,.
\label{eq9.54}
\end{eqnarray}
The amplitudes  $r_{L,R}$ and $t_{L,R}$ indicate the as--yet unknown strengths
of the LCP and RCP  components of
the reflected and transmitted plane waves, both of which are
elliptically polarized in general. 

As the tangential components of $\bf E$ and $\bf H$ must be
continuous across the planes $z=0$ and $z=L$, 
the boundary
values $\bar{\psi}(0)$
and $\bar{\psi}(L)$ 
can be fixed by virtue of \r{eq9.50}--\r{eq9.54}. Hence,
\begin{eqnarray}
\label{eq9.60}
&&
\bar{\psi}(0) 
=\frac{1}{\sqrt{2}}\,\bar{K}\cdot
\les\begin{array}{c}i(\aL-\aR)\\ -(\aL+\aR)\\ 
-i(\rL-\rR) \\ \rL+\rR \end{array}\ris\,,
\end{eqnarray}
and
\begin{eqnarray}
\label{eq9.61}
&&
\bar{\psi}(L)
=\frac{1}{\sqrt{2}}\,\bar{K}\cdot
\les\begin{array}{c}i(\tL-\tR)\\ -(\tL+\tR)\\ 
0 \\ 0\end{array}\ris\,,
\end{eqnarray}
where  

\begin{equation}
\bar{K}
=
\les\begin{array}{cccc}
-\sphi & -\cphi\,\ctheta & -\sphi & \cphi\,\ctheta\\[2pt]
\cphi & -\sphi\,\ctheta & \cphi & \sphi\,\ctheta\\[2pt]
-\etao^{-1}\,\cphi\,\ctheta & \etao^{-1}\,\sphi &\etao^{-1}\,\cphi\,\ctheta & \etao^{-1}\,\sphi\\[2pt]
-\etao^{-1}\,\sphi\,\ctheta &-\etao^{-1}\,\cphi & \etao^{-1}\,\sphi\,\ctheta & -\etao^{-1}\,\cphi
\end{array}\ris\,.
\label{eq7.39}
\end{equation}

The reflection--transmission problem thus amounts to
four simultaneous, linear algebraic equation stated
in matrix form as
\begin{equation}
\label{eq9.63}
\les\begin{array}{c}i(\tL-\tR)\\ -(\tL+\tR)\\ 
0 \\ 0\end{array}\ris=
\bar{K}^{-1}\cdot
\bar{U}(L)\cdot
\bar{K}\cdot
\les\begin{array}{c}i(\aL-\aR)\\ -(\aL+\aR)\\ 
-i(\rL-\rR) \\ \rL+\rR \end{array}\ris\,.
\end{equation}
This set of equations can be solved by standard matrix 
manipulations
to compute the reflection and transmission coefficients.

It is usually convenient
to define reflection and transmission coefficients. 
These appear as the elements of the 2$\times$2 matrixes in the following
relations:
\begin{equation}
\label{eq9.55}
\les\begin{array}{c}\rL\\\rR\end{array}\ris
=
\les\begin{array}{cc}\rLL & \rLR\\\rRL & \rRR\end{array}\ris
\,
\les\begin{array}{c}\aL\\\aR\end{array}\ris\,,
\end{equation}
\begin{equation}
\label{eq9.56}
\les\begin{array}{c}\tL\\\tR\end{array}\ris
=
\les\begin{array}{cc}\tLL & \tLR\\\tRL & \tRR\end{array}\ris
\,
\les\begin{array}{c}\aL\\\aR\end{array}\ris\,.
\end{equation}
Co--polarized coefficients have both subscripts identical, but
cross--polarized coefficients do not. The square of the magnitude
of a reflection or transmission coefficient is the corresponding
reflectance or transmittance;  thus, $\RLR = \vert\rLR\vert^2$ is
the reflectance corresponding to the reflection coefficient $\rLR$,
and so on.
The principle of conservation of energy mandates
\index{energy conservation}
the constraints
\begin{equation}
\label{eq9.59}
\left.
\begin{array}{l}
\RLL + \RRL + \TLL + \TRL \leq 1\\[2pt]
\RRR + \RLR + \TRR + \TLR \leq 1
\end{array}\ric\,,
\end{equation}
the inequalities turning to equalities only in the
absence of dissipation inside the SCM slab.

\subsection{Normal incidence}\label{axprop}
For normal incidence, electromagnetic wave propagation in the SCM occurs
parallel to the axis
of structural chirality, and a special case amenable to algebraic
analysis emerges \cite{LW95}.
Then $\kappa = 0$, and (\ref{oblique}) simplifies to 
\begin{equation}  \label{axial1}
\frac{d}{dz}{\bar{\psi}}^{\prime }(z)= i \bar{A}^\prime_{ax}\cdot{\bar{\psi}}%
^{\prime }(z)\,, \qquad 0 < z <L\,,
\end{equation}
wherein the matrix 
\begin{equation}  \label{axial2}
\bar{A}^\prime_{ax} = \left( 
\begin{array}{cccc}
0 & -i\frac{h\pi}{\Omega} & 0 & \omega\mu_o \\ 
i\frac{h\pi}{\Omega} & 0 & -\omega\mu_o & 0 \\ 
-\omega\epsilon_o\epsilon_E & -\omega\epsilon_o\epsilon_B & 0 & -i\frac{h\pi%
}{\Omega} \\ 
\omega\epsilon_o\epsilon_D & \omega\epsilon_o\epsilon_E & i\frac{h\pi}{\Omega%
} & 0
\end{array}
\right) \,
\end{equation}
is independent of $z$, and 
\begin{eqnarray}
&& \epsilon_B=\epsilon_2^{(0)} -\frac{\epsilon_2^{(0)}}{\epsilon_1^{(0)}}
\,\epsilon_m\,, \\[5pt]
&& \epsilon_D=\epsilon_d -\frac{\epsilon_2^{(0)}}{\epsilon_1^{(0)}} \left[
\epsilon_\iota\cos\chi+(\epsilon_j+\epsilon_\ell) \frac{\sin 2\chi}{2}%
+\epsilon_k\sin\chi \right]\,, \\[5pt]
&&\epsilon_E= \frac{\epsilon_2^{(0)}}{\epsilon_1^{(0)}}\,
(\epsilon_e+\epsilon_h)\,.
\end{eqnarray}
The solution of (\ref{axial1}) therefore is straightfoward:  
\begin{equation}  \label{axial3}
\bar{U}^\prime(z)=
\exp\left[ i z\bar{A}^\prime_{ax}\right]
\end{equation}

But an even more illuminating solution becomes available by further
extending the Oseen transformation \cite{genOs}. Let us define the column vector 
\begin{equation}
{\bar{\psi}}^{{\prime\prime} }(z)=\bar{M}\left(h\xi\right)\cdot {\bar{\psi}}%
^\prime(z)\,,
\end{equation}
where 
\begin{equation}
\xi= \frac{1}{2}\tan^{-1}\left(\frac{2h\epsilon_E}{\epsilon_D-\epsilon_B}%
\right)\,.
\end{equation}
Then, (\ref{axial1}) transforms to 
\begin{equation}  \label{axial4}
\frac{d}{dz}{\bar{\psi}}^{{\prime\prime} }(z)= i \bar{A}^{{\prime\prime}%
}_{ax}\cdot{\bar{\psi}}^{{\prime\prime} }(z)\,, \qquad 0 < z <L\,,
\end{equation}
where 
\begin{equation}  \label{axial5}
\bar{A}^\dprime_{ax} = \left( 
\begin{array}{cccc}
0 & -i\frac{h\pi}{\Omega} & 0 & \omega\mu_o \\ 
i\frac{h\pi}{\Omega} & 0 & -\omega\mu_o & 0 \\ 
0 & -\omega\epsilon_o\epsilon_{B\xi} & 0 & -i\frac{h\pi}{\Omega} \\ 
\omega\epsilon_o\epsilon_{D\xi} & 0 & i\frac{h\pi}{\Omega} & 0
\end{array}
\right) \,,
\end{equation}
\begin{eqnarray}
\label{triple1}
&& \epsilon_{B\xi}=\frac{1}{2}\left[ \epsilon_B +\epsilon_D + \frac{%
\left(\epsilon_B -\epsilon_D\right)^2+4 \epsilon_E^2 }{\epsilon_B -\epsilon_D%
}\,\cos 2\xi\right]\,, \\[5pt]
\label{triple2}
&&\epsilon_{D\xi}=\frac{1}{2}\left[ \epsilon_B +\epsilon_D - \frac{%
\left(\epsilon_B -\epsilon_D\right)^2+4 \epsilon_E^2 }{\epsilon_B -\epsilon_D%
}\,\cos 2\xi\right]\,.
\end{eqnarray}

The lower left quadrant of $\bar{A}^\dprime_{ax}$ is antidiagonal; so is the
lower left quadrant of $\bar{A}^\prime_{ax}$ when $E_z^{dc}=0$. Thus, by
comparison to extant results for non--electro--optic SCMs \cite{LakhtakiaB},
we can state that the center--wavelength of the Bragg regime for normal
incidence is 
\begin{equation}  \label{Br-def}
\lambdao^{Br} = \Omega \left( \sqrt{\epsilon_{B\xi}}+ \sqrt{\epsilon_{D\xi}}%
\right)\,
\end{equation}
and the corresponding full--width--at--half--maximum (FWHM) bandwidth is 
\begin{equation}  \label{dBr-def}
(\Delta\lambda_o)^{Br}=2\Omega \Big\vert\sqrt{\epsilon_{B\xi}}- \sqrt{%
\epsilon_{D\xi}}\Big\vert\,,
\end{equation}
with the assumption that dissipation in the SCM is negligibly small and
dispersion in the constitutive properties can be ignored \cite{VLepsa}.

Correct to the second order in terms such as $r_{41}E_z^{dc}$, we get 
\begin{eqnarray}  \label{dd2a}
&&\epsilon_{B\xi}^{1/2}\approx \sqrt{\epsilon_2^{(0)}} \left[ 1 -\frac{1}{2}%
\frac{\epsilon_m}{\epsilon_1^{(0)}} -\frac{1}{8}\left(\frac{\epsilon_m}{%
\epsilon_1^{(0)}}\right)^2 +\frac{1}{2}\left(\frac{\epsilon_2^{(0)}}{%
\epsilon_1^{(0)}}\right)^2 \frac{(\epsilon_e+\epsilon_h)^2}{%
\epsilon_2^{(0)}(\epsilon_2^{(0)}-\epsilon_d)}\right]\,, \\
\label{dd2b}
&&\epsilon_{D\xi}^{1/2}\approx \sqrt{\epsilon_d} \left[ 1 +\frac{1}{2}\frac{%
\epsilon_D-\epsilon_d}{\epsilon_d} -\frac{1}{8}\left(\frac{%
\epsilon_D-\epsilon_d}{\epsilon_d}\right)^2 -\frac{1}{2}\left(\frac{%
\epsilon_2^{(0)}}{\epsilon_1^{(0)}}\right)^2 \frac{(\epsilon_e+\epsilon_h)^2%
}{\epsilon_d(\epsilon_2^{(0)}-\epsilon_d)}\right]\,,
\end{eqnarray}
which allows the delineation of the effect of the {\it local} crystallographic
classification (as captured by the various electro--optic coefficients $r_{JK}$) on the extent of
the Bragg regime. As an example, the foregoing expressions may be set down
as 
\begin{eqnarray}\label{cdd1a}
&&\epsilon_{B\xi}^{1/2}\approx \sqrt{\epsilon_2^{(0)}} \left[ 1 -\frac{1}{2}%
\epsilon_2^{(0)}E_z^{dc} r_{21}-\frac{1}{8}\left(\epsilon_2^{(0)}E_z^{dc}
r_{21}\right)^2 +\frac{1}{2}\frac{\epsilon_2^{(0)}}{\epsilon_2^{(0)}-%
\epsilon_3^{(0)}} \left(\epsilon_3^{(0)}E_z^{dc} r_{41}\right)^2\right]\,, \\
\label{cdd1b}
&&\epsilon_{D\xi}^{1/2}\approx \sqrt{\epsilon_3^{(0)}} \left[ 1 -\frac{1}{2}%
\epsilon_3^{(0)}E_z^{dc} r_{31} -\frac{1}{8} \left(\epsilon_3^{(0)}E_z^{dc}
r_{31}\right)^2 -\frac{1}{2}\frac{\epsilon_3^{(0)}}{\epsilon_2^{(0)}-%
\epsilon_3^{(0)}} \left(\epsilon_2^{(0)}E_z^{dc} r_{41}\right)^2\right]\,,
\end{eqnarray}
when $\chi=0$. As another example, we get 
\begin{eqnarray}  \label{cdd2a}
&&\epsilon_{B\xi}^{1/2}\approx \sqrt{\epsilon_2^{(0)}} \left[ 1 - \frac{1}{2}%
\epsilon_2^{(0)}E_z^{dc} r_{23} -\frac{1}{8}\left(\epsilon_2^{(0)}E_z^{dc}
r_{23}\right)^2 +\frac{1}{2}\frac{\epsilon_2^{(0)}}{\epsilon_2^{(0)}-%
\epsilon_1^{(0)}} \left(\epsilon_1^{(0)}E_z^{dc} r_{63}\right)^2\right]\,, \\
\label{cdd2b}
&&\epsilon_{D\xi}^{1/2}\approx \sqrt{\epsilon_1^{(0)}} \left[ 1 - \frac{1}{2}%
\epsilon_1^{(0)}E_z^{dc} r_{13} -\frac{1}{8}\left(\epsilon_1^{(0)}E_z^{dc}
r_{13}\right)^2 -\frac{1}{2}\frac{\epsilon_1^{(0)}}{\epsilon_2^{(0)}-%
\epsilon_1^{(0)}} \left(\epsilon_1^{(0)}E_z^{dc} r_{63}\right)^2\right]\,,
\end{eqnarray}
when $\chi=\pi/2$.

Equations (\ref{cdd2a}) and (\ref{cdd2b}) do not hold for locally uniaxial
SCMs (i.e., when $\epsilon_1^{(0)} = \epsilon_2^{(0)}$) \cite{RL06}. Fresh
analysis reveals 
\begin{eqnarray}
&&\epsilon_{B\xi}^{1/2}\approx \sqrt{\epsilon_1^{(0)}} \left[ 1 + \frac{1}{4}
\epsilon_1^{(0)}E_z^{dc}\left[\beta(r_{13}-r_{23})-(r_{13}+r_{23})\right]%
\right.  \nonumber \\[5pt]
&&\left.\quad- \frac{1}{32} \left\{\epsilon_1^{(0)}E_z^{dc}\left[%
\beta(r_{13}-r_{23})-(r_{13}+r_{23})\right]\right\}^2 \right]\,,
\label{cdd3} \\[9pt]
&&\epsilon_{D\xi}^{1/2}\approx \sqrt{\epsilon_1^{(0)}} \left[ 1 - \frac{1}{4}
\epsilon_1^{(0)}E_z^{dc}\left[\beta(r_{13}-r_{23})+(r_{13}+r_{23})\right]%
\right.  \nonumber \\[5pt]
&&\left.\quad- \frac{1}{32} \left\{\epsilon_1^{(0)}E_z^{dc}\left[%
\beta(r_{13}-r_{23})+(r_{13}+r_{23})\right]\right\}^2 \right]\,,
\label{cdd4}
\end{eqnarray}
when $\chi=\pi/2$ and $\epsilon_2^{(0)} = \epsilon_1^{(0)}$; here, 
\begin{equation}  \label{cdd5}
\beta=\left[ 1+ \left(\frac{2 r_{63}}{r_{13}-r_{23}}\right)^2\right]^{1/2}\,.
\end{equation}
Accordingly, the FWHM bandwidth of the Bragg regime turns out to be
proportional to the magnitude of the dc electric field as per 

\begin{equation}  \label{cdd6}
(\Delta\lambda_o)^{Br}=2\Omega (\epsilon_1^{(0)})^{3/2}\Big\vert\beta%
E_z^{dc} (r_{13}-r_{23})\left[ 1+\frac{1}{4}%
\epsilon_1^{(0)}E_z^{dc}(r_{13}+r_{23})\right]\Big\vert\,.
\end{equation}

Equation (\ref{cdd6}) indicates that the Bragg regime vanishes for normal
incidence   on a locally uniaxial, non--electro--optic
SCM with $\chi =\pi/2$ \cite{VLprsa1998}, but can be generated by the
appropriate application of a dc electric field if the SCM is electro--optic.
This is the generalization of a result previously obtained for SCMs with
local $\bar{4}2m$ point group symmetry \cite{RL06}. Obviously, this
conclusion may be exploited for optical switching applications for turning
on or off a chosen circular polarization state.

The foregoing statements may be applied to locally biaxial SCMs as well in
the following manner. Suppose that 
\begin{equation}  \label{dd3}
\chi= \tan^{-1}\left[\left( \frac{\epsilon_1^{(0)}}{\epsilon_3^{(0)}}%
\right)\, \left(\frac{\epsilon_2^{(0)}-\epsilon_3^{(0)}}{\epsilon_1^{(0)}-%
\epsilon_2^{(0)}}\right)\right]^{1/2}\,;
\end{equation}
then, $\epsilon_d=\epsilon_2^{(0)}$ by virtue of (\ref{epsd-def}). For
non--electro--optic SCMs, (\ref{dd3}) defines the \emph{pseudoisotropic point%
} \cite{pse1,pse2}: the Bragg regime for normal incidence vanishes, as may
be seen by substituting (\ref{triple1}) and (\ref{triple2}) in (\ref{dBr-def}),
setting $E_z^{dc}=0$ in the resulting expression, and making use of (\ref
{dd3}) thereafter. However, the Bragg regime can be restored by the
application of $E_z^{dc}$, provided the SCM is electro--optic; the
bandwidth of the Bragg regime can thus be electrically controlled. A general
expression for $(\Delta\lambda_o)^{Br}$ at the pseudoisotropic point is far
too cumbersome for reproduction here.

\section{Numerical results and discussion}\label{nrd}

A Mathematica program was written to compute
the reflectances and transmittances of a SCM slab of thickness $L$,
on which an arbitrarily polarized plane wave is incident from the half--space
$z<0$ with an arbitrarily oriented wave vector. The principle of
conservation of energy was verified to within $\pm0.1\%$ error for
all results presented in this section. All calculations were made
for structurally right--handed SCMs.

\subsection{Locally isotropic SCMs}
Locally isotropic SCMs are characterized by $\epsilon_1^{(0)} = \epsilon_2^{(0)}= \epsilon_3^{(0)}$, and therefore cannot evince the circular Bragg phenomenon 
in the absence of a dc electric field. However, electro--optic materials
of crystallographic classes $\bar{4}3m$ and $23$ have $r_{41}=r_{52}=r_{63}\ne 0$
with all other $r_{JK}\equiv 0$ \cite[pp. 170--176]{LB},
and can therefore exhibit CBP when $\Ezdc\ne 0$ \c{Lmotl2006}.

Analysis of $\bar{A}^\prime_{ax}$ reveals that the application of $\Ezdc$ would
be infructous towards the exhibition of the CBP
for normal incidence (i.e., $\theta=0^\circ$), if $\chi=45^\circ$; but it would be most (and
equally) effective for $\chi=0^\circ$ and $\chi=90^\circ$. The analytic continuability
of $\bar{A}^\prime(z)$ with respect to $\kappa$ suggests that the foregoing statement would be substantially true even for oblique incidence, at least for small and moderate values of $\theta$; and the validity of the suggestion was confirmed computationally.

\begin{figure}[!ht]
\centering \psfull
\epsfig{file=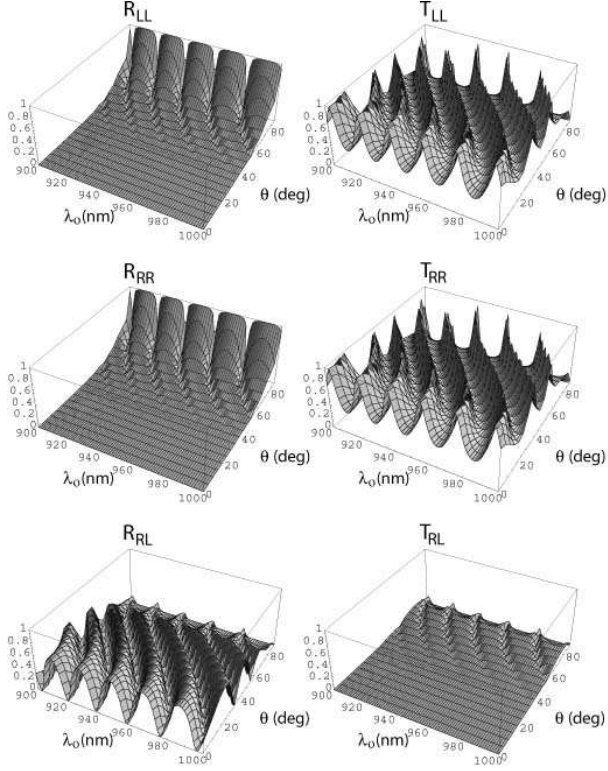,width=8cm }
\caption{Reflectances and transmittances of a locally isotropic SCM slab of thickness
$L=50\,\Omega$ as functions of the free--space wavelength $\lambda_o$
and the incidence angle $\theta$, when $\Ezdc=0$ and $\phi=0^\circ$. The local crystallographic
class 
of the SCM is $\bar{4}3m$. Other parameters are: $\epsilon_1^{(0)} = \epsilon_2^{(0)}= \epsilon_3^{(0)}=8.94$, $r_{41}=r_{52}=r_{63}=4.04
\times 10^{-12}$~m~V$^{-1}$, all other $r_{JK}=0$,
$h=1$, $\Omega=160$~nm, and $\chi=90^\circ$. These plots are   the same as for $\chi=0^\circ$. As   $T_{LR}=T_{RL}$   and $R_{LR}=R_{RL}$ to numerical accuracy, the plots of  $T_{LR}$ and
 $T_{LR}$ are not shown.
}
\label{bar43m-1}
\end{figure}

\begin{figure}[!ht]
\centering \psfull
\epsfig{file=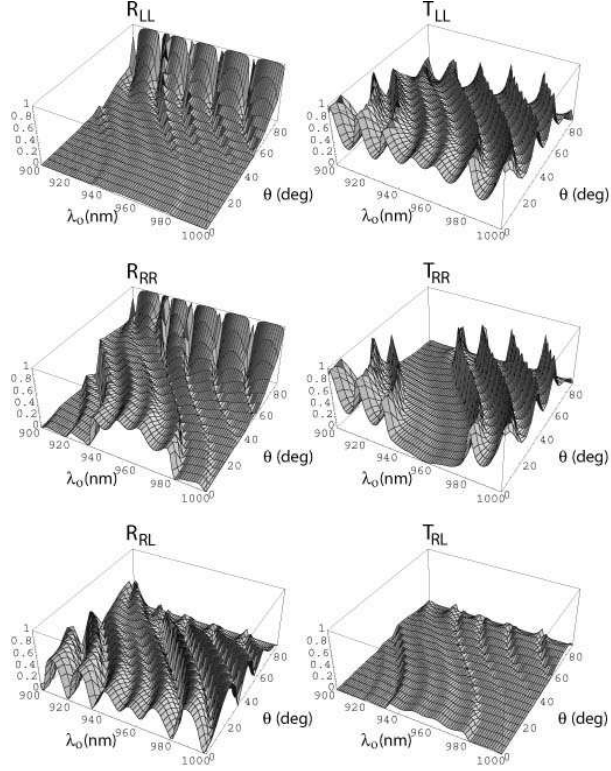,width=8cm }
\caption{Same as Fig. \ref{bar43m-1}, except that a  dc voltage $V_{dc}= 8$ kV is applied
between the planes $z=0$ and $z=L$; thus, $\Ezdc= V_{dc}/L= 1$~GV~m$^{-1}$.  As the differences between $T_{LR}$ and $T_{RL}$ are very
small, and $R_{LR}=R_{RL}$ to numerical accuracy, the plots of  $T_{LR}$ and
 $T_{LR}$ are not shown.
Note that $r_{41}\Ezdc=0.00404$ is much smaller
than $1/\epsilon_1^{(1)}=0.1118$.
}
\label{bar43m-2}
\end{figure}

Figures \ref{bar43m-1} and \ref{bar43m-2} help elucidate the effect of the
dc electric field. The first figure shows the reflectance and transmittance
spectrums for  the incidence angle $\theta\in\left[0^\circ,90^\circ\right)$ when $\Ezdc=0$
and $\phi=0^\circ$. The second figure has the same spectrums but when a dc voltage $V_{dc}=8$~kV
is applied across the planes $z=0$ and $z=L$. The structurally right--handed
SCM slab was taken to be
$50\Omega$ thick, $\Omega=160$~nm, $\chi=90^\circ$, and its local relative permittivity 
matrix and electro--optic parameters were chosen to be the same as of
zinc telluride \cite[p. 173]{LB}. 

In a $\sim40$--nm regime centered at $\lambdao^{Br}=956$~nm, $R_{RR}$ is high and $T_{RR}$ is low when $\theta=0^\circ$, in Fig. \ref{bar43m-2}. This is the Bragg regime, which
blueshifts as $\cos\theta$ decreases from unity in magnitude.
The Bragg regime is entirely absent in Fig. \ref{bar43m-1} when no dc voltage is applied. The Bragg regime is also absent in Fig. \ref{bar43m-2}, even when the dc voltage is applied, for incident LCP plane waves. Clearly therefore, the Pockels effect has engendered the CBP in a SCM with a local crystallographic symmetry that is isotropic. 

The cross--polarized reflectances
and transmittances in the Bragg regime can be reduced by a variety of
 impedance--matching techniques \cite{HWAML}, and  thereafter the SCM slab can function
 as an electrically switchable circular--polarization rejection filter for incident plane waves
 of the same handedness as the SCM.

%%%%%%%%%%  Figure 1 begins %%%%%%%%%%%%
%%%%%%%%%%  Figure 1 ends  %%%%%%%%%%%%

%%%%%%%%%%  Figure 2 begins %%%%%%%%%%%%
%%%%%%%%%%  Figure 2 ends  %%%%%%%%%%%%

%%%%%%%%%%  Figure 3 begins %%%%%%%%%%%%
\begin{figure}[!ht]
\centering \psfull
\epsfig{file=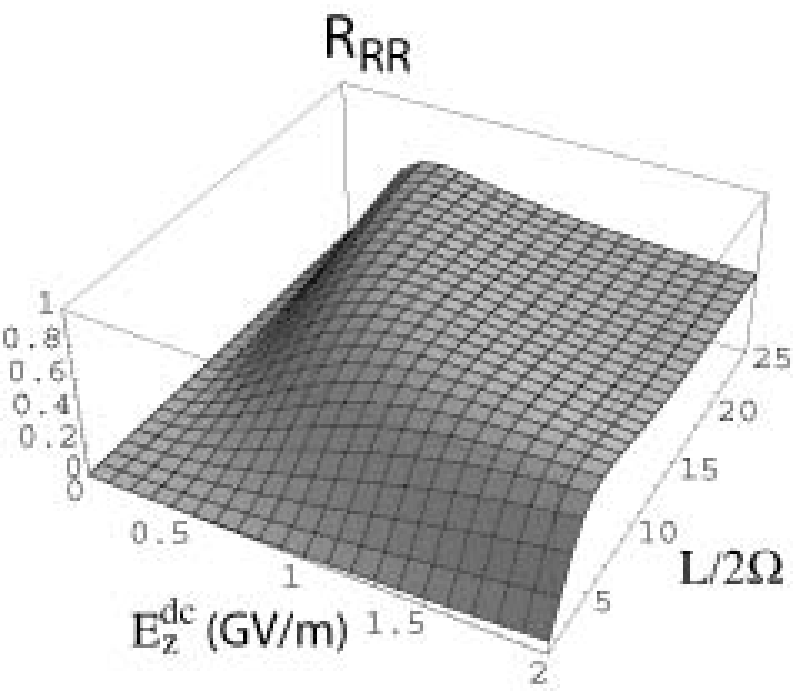,width=4cm }
\caption{Reflectance $R_{RR}$ of  a locally isotropic SCM slab as a function
of $L/2\Omega$ and $\Ezdc$. The local crystallographic
class of the SCM is $\bar{4}3m$, with   $\epsilon_1^{(0)} = \epsilon_2^{(0)}= \epsilon_3^{(0)}=8.94$, $r_{41}=r_{52}=r_{63}=4.04
\times 10^{-12}$~m~V$^{-1}$, all other $r_{JK}=0$,
$h=1$, $\Omega=160$~nm, and $\chi=90^\circ$. The
angles of incidence $\theta=\phi=0^\circ$, and the wavelength $\lambdao=956$~nm
lies in the middle of the Bragg regime for normal incidence.  }
\label{bar43m-3}
\end{figure}

%%%%%%%%%%  Figure 3 ends  %%%%%%%%%%%%

 It is known from many studies on chiral sculptured thin films
 as well as cholesteric liquid crystals that the CBP first
 deepens and then saturates, as the normalized thickness $L/\Omega$ increases \cite{LakhtakiaB,StJ}. The well--developed CBP manifests itself as a feature with
 a tall top--hat profile in the spectrum of $R_{RR}$ (resp. $R_{LL}$)
 for normal and near--normal  incidence on a structurally right--handed
 (resp. left--handed) slab.
 Further increase of thickness beyond a certain value of $L/\Omega$ is
 therefore infructous. The same conclusion should hold true for an electro--optic SCM if $\Ezdc$ were held fixed,
 and is indeed borne out in Fig.~\ref{bar43m-3} by the plot of $R_{RR}$ for a structurally
 right--handed SCM slab with other parameters the same as for the previous
 figure. This plot holds at the center--wavelength of the Bragg regime for
 normal incidence.
 Figure~\ref{bar43m-3} also suggests that after a certain value, even an
 increase in $\Ezdc$ for a fixed $L/2\Omega$ would lead to diminishing returns, if the objective is to
 maximize $R_{RR}$. 
 
 Electrical control of SCMs appears to require high dc voltages. These can be comparable
with the half--wave voltages of electro--optic materials \cite[p. 420]{YY6}, 
which are often in the 1--10~kV range. We
must also note that the required magnitudes of
$\Ezdc$ are much smaller than
the characteristic atomic electric field strength \cite[p. 3]{Boyd}. 
Similarly
high voltages are often applied to electro--optic films, albeit to create
electric fields that are two orders--of--magnitude smaller
than $\Ezdc$ in Fig.~\ref{bar43m-2} \cite{Scry}. The
possibility of electric breakdown  exists, but it would significantly
depend on the time that the dc voltage would be switched on for.

\subsection{Locally uniaxial SCMs}\label{uni}
Locally uniaxial SCMs are characterized by  $\epsilon_1^{(0)} = \epsilon_2^{(0)}\ne
 \epsilon_3^{(0)}$. Crystals in 13
classes divided into the trigonal, tetragonal, and hexagonal families can exhibit the Pockels effect
\cite{Boyd}. Lithium niobate and potassium dihydrogen phosphate are
perhaps the most well--known uniaxial electro--optic materials, but a host of other
materials with similar properties also exist \cite[pp. 176--201]{LB}.

\begin{figure}[!ht]
\centering \psfull
\epsfig{file=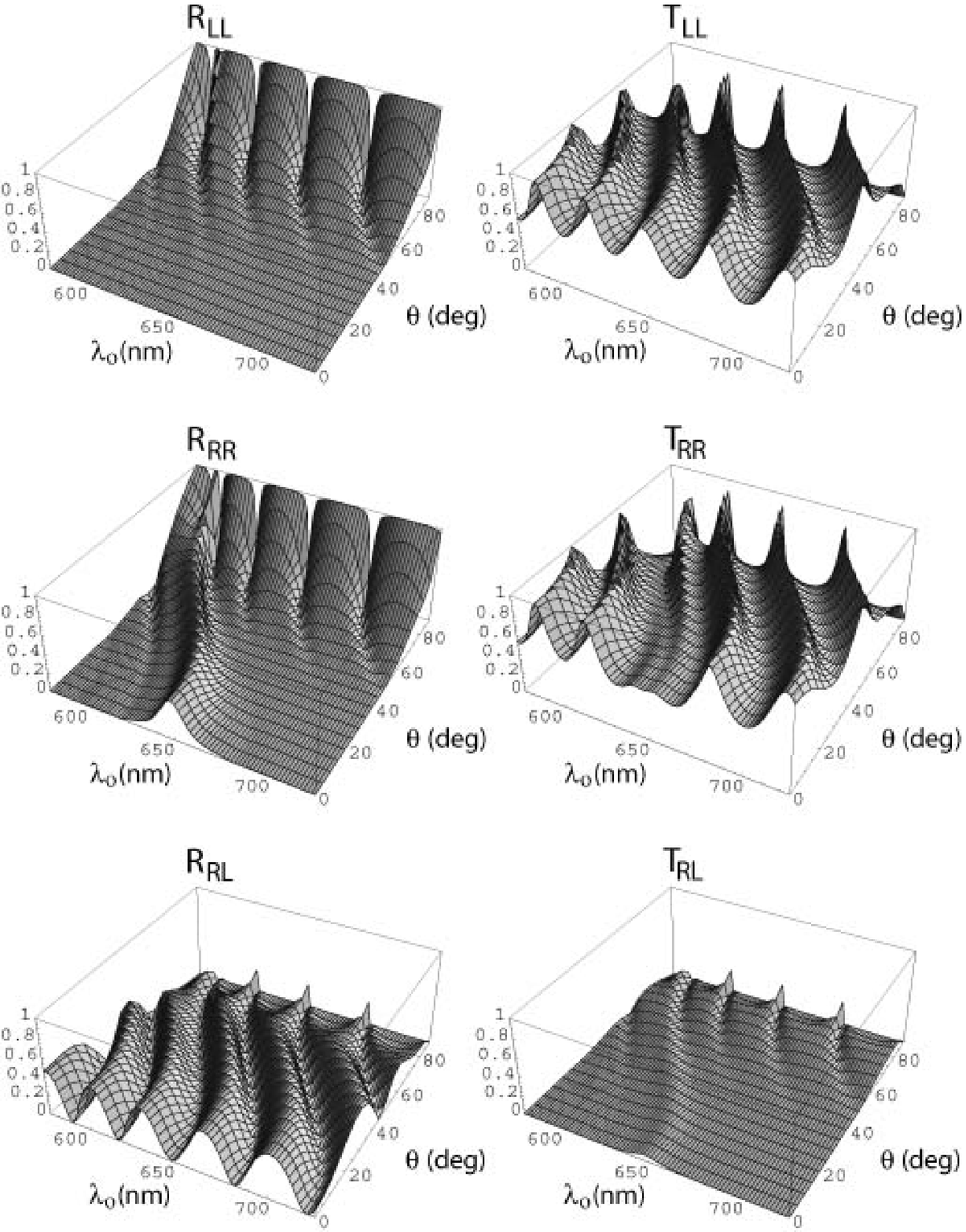,width=8cm }
\caption{Reflectances and transmittances of a locally uniaxial SCM slab of thickness
$L=20\,\Omega$ as functions of the free--space wavelength $\lambda_o$
and the incidence angle $\theta$, when $\Ezdc=0$ and $\phi=0^\circ$. The local crystallographic
class 
of the SCM is trigonal $3m$. Other parameters are: $\epsilon_1^{(0)} = \epsilon_2^{(0)}= 5.48$,$\epsilon_3^{(0)}=5.04$, 
$r_{22}=-r_{12}=-r_{61}=6.8
\times 10^{-12}$~m~V$^{-1}$, $r_{13}=r_{23}=9.6
\times 10^{-12}$~m~V$^{-1}$, $r_{33}=30.9\times 10^{-12}$~m~V$^{-1}$,
$r_{42}=r_{51}=32.6\times 10^{-12}$~m~V$^{-1}$, all other $r_{JK}=0$,
$h=1$, $\Omega=140$~nm, and $\chi=45^\circ$. As   $T_{LR}=T_{RL}$   and $R_{LR}=R_{RL}$ to numerical accuracy, the plots of  $T_{LR}$ and
 $T_{LR}$ are not shown.}
\label{trigonal3m-1}
\end{figure}

Locally uniaxial SCMs should exhibit the CBP even in the absence of a
dc electric field, and indeed they do, as is evident from Fig.~\ref{trigonal3m-1} 
which shows the reflectance and transmittance
spectrums for  incidence angles $\theta\in\left[0^\circ,90^\circ\right)$
and $\phi=0^\circ$ when $\Ezdc=0$. The chosen SCM has trigonal $3m$
as its local crystallographic class, with the values of the relative permittivity scalars and the electro--optic coefficients the
same as for lithium niobate \cite[p. 184]{LB}. The plots of $R_{RR}$ 
and $T_{RR}$ show the Bragg regime
centered about $\lambdao^{Br}=648$~nm when $\theta=0^\circ$, the Bragg regime
exhibiting a blueshift with decrease of $\cos\theta$. The SCM slab is not very
thick ($L=20\Omega$); hence, the CBP is not fully developed \cite{LakhtakiaB,StJ}.

\begin{figure}[!ht]
\centering \psfull
\epsfig{file=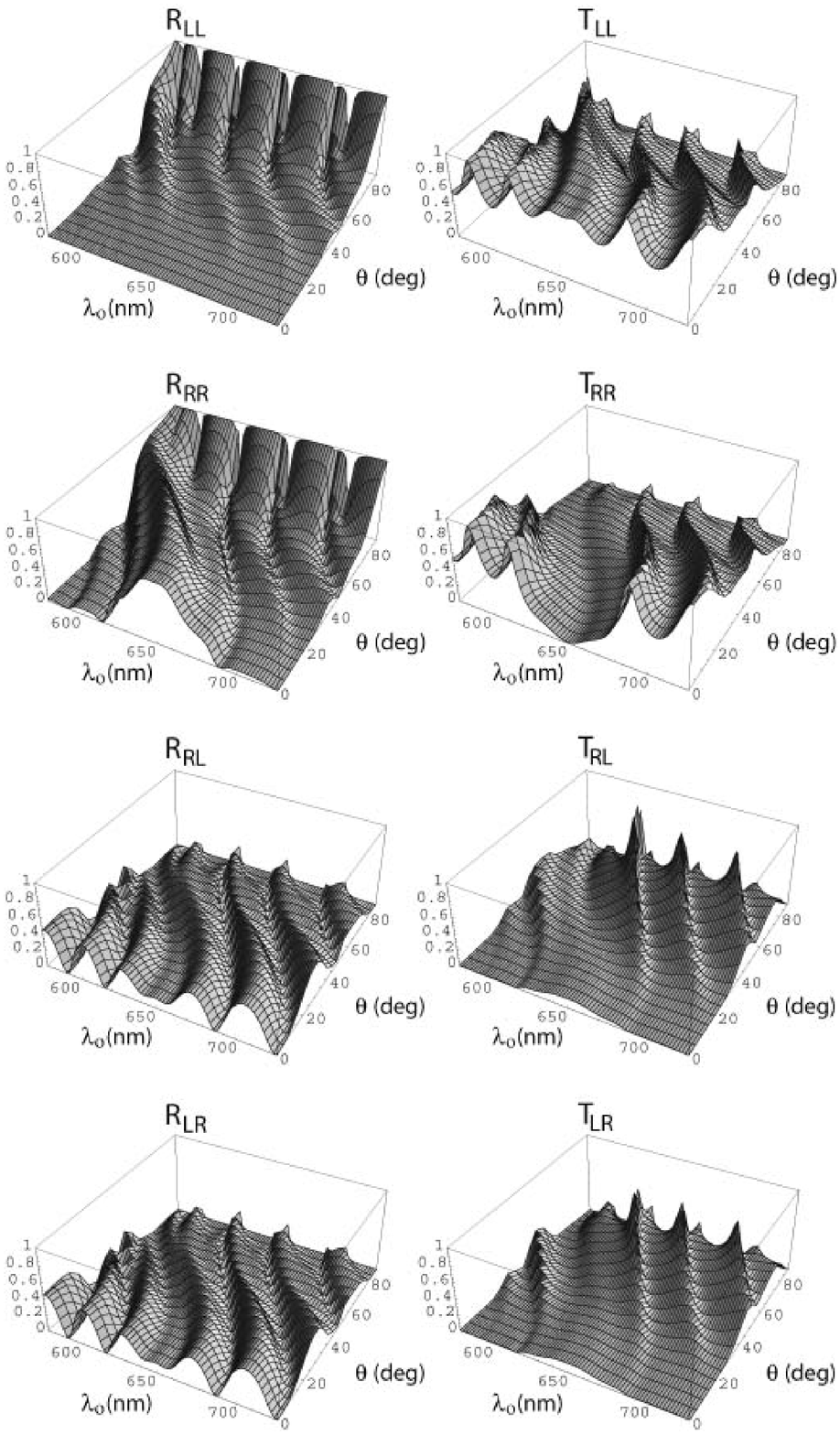,width=8cm }
\caption{Same as Fig. \ref{trigonal3m-1}, except that a  dc voltage $V_{dc}= 5$~kV is applied
between the planes $z=0$ and $z=L$; thus, $\Ezdc= V_{dc}/L= 1.79$~GV~m$^{-1}$.
}
\label{trigonal3m-2}
\end{figure}

Figure~\ref{trigonal3m-2} has the same reflectance and transmittance plots as the
preceding figure, except that a dc voltage of 5~kV is applied across
the SCM slab. The CBP in Fig.~\ref{trigonal3m-2} is definitely enhanced
in comparison to Fig.~\ref{trigonal3m-1}; calculated results not presented here indicate
even a better developed CBP in the form of a broad top--hat profile of
the $R_{RR}$--ridge for higher values of $\Ezdc$.  

\begin{figure}[!ht]
\centering \psfull
\epsfig{file=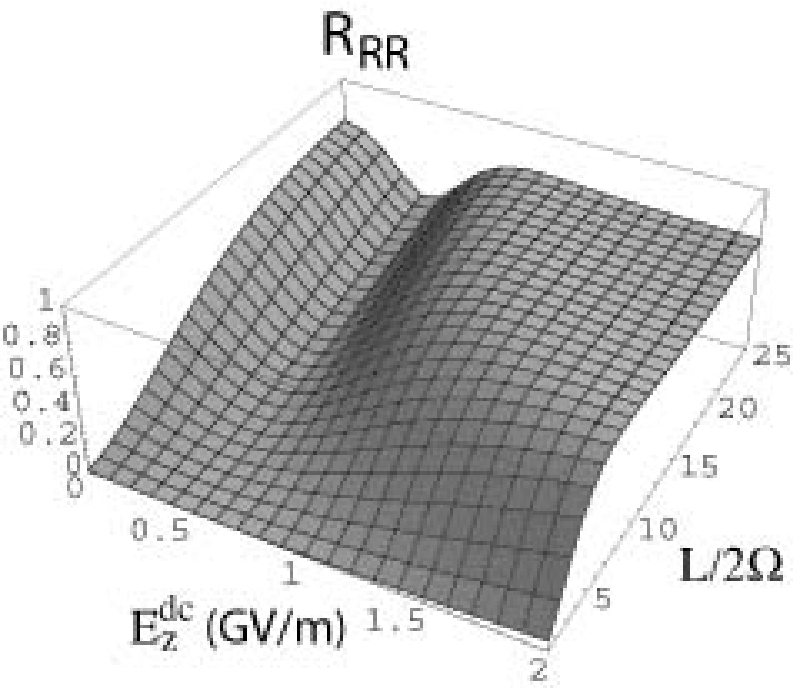,width=4cm }
\caption{Reflectance $R_{RR}$ of  a locally uniaxial SCM slab as a function
of $L/2\Omega$ and $\Ezdc$. The local crystallographic
class 
of the SCM is trigonal $3m$, with $\epsilon_1^{(0)} = \epsilon_2^{(0)}= 5.48$,$\epsilon_3^{(0)}=5.04$, 
$r_{22}=-r_{12}=-r_{61}=6.8
\times 10^{-12}$~m~V$^{-1}$, $r_{13}=r_{23}=9.6
\times 10^{-12}$~m~V$^{-1}$, $r_{33}=30.9\times 10^{-12}$~m~V$^{-1}$,
$r_{42}=r_{51}=32.6\times 10^{-12}$~m~V$^{-1}$, all other $r_{JK}=0$,
$h=1$, $\Omega=140$~nm, and $\chi=45^\circ$.
The
angles of incidence $\theta=\phi=0^\circ$, and the wavelength $\lambdao=648$~nm
lies in the middle of the Bragg regime for normal incidence.  }
\label{trigonal3m-3}
\end{figure}

Thus, there are two ways to enhance the CBP for exploitation 
in circular--polarization rejection
filters. The first is to use thicker SCM slabs, i.e., the ratios $L/2\Omega$ are large.
The second is to use higher $V_{dc}$. The interplay between these two factors
is indicated in Fig.~\ref{trigonal3m-3}, wherein $R_{RR}$ for normal
incidence and $\lambdao=648$~nm is plotted as a function of $L/2\Omega$
and $\Ezdc$. This figure clearly indicates that the exploitation of
the Pockels effect will lead to thinner filters. 

Qualitatively comparable results were obtained when the local crystallographic class
was changed from trigonal $3m$ to any of the other 12 trigonal, tetragonal or
hexagonal classes.

%%%%%%%%%%  Figure 4 begins %%%%%%%%%%%%
%%%%%%%%%%  Figure 4 ends  %%%%%%%%%%%%

%%%%%%%%%%  Figure 5 begins %%%%%%%%%%%%
%%%%%%%%%%  Figure 5 ends  %%%%%%%%%%%%

%%%%%%%%%%  Figure 6 begins %%%%%%%%%%%%
%%%%%%%%%%  Figure 6 ends  %%%%%%%%%%%%

\subsection{Locally biaxial SCMs}\label{bi}

Locally biaxial SCMs are characterized by  $\epsilon_1^{(0)} \ne \epsilon_2^{(0)}\ne
 \epsilon_3^{(0)}$. Crystals in 5
classes separated into the orthorhombic, monoclinic, and triclinic families can exhibit the Pockels effect
\cite{Boyd}. Potassium niobate and sodium barium niobate are
well--known biaxial electro--optic materials \cite{SDG,ZSB}.

\begin{figure}[!ht]
\centering \psfull
\epsfig{file=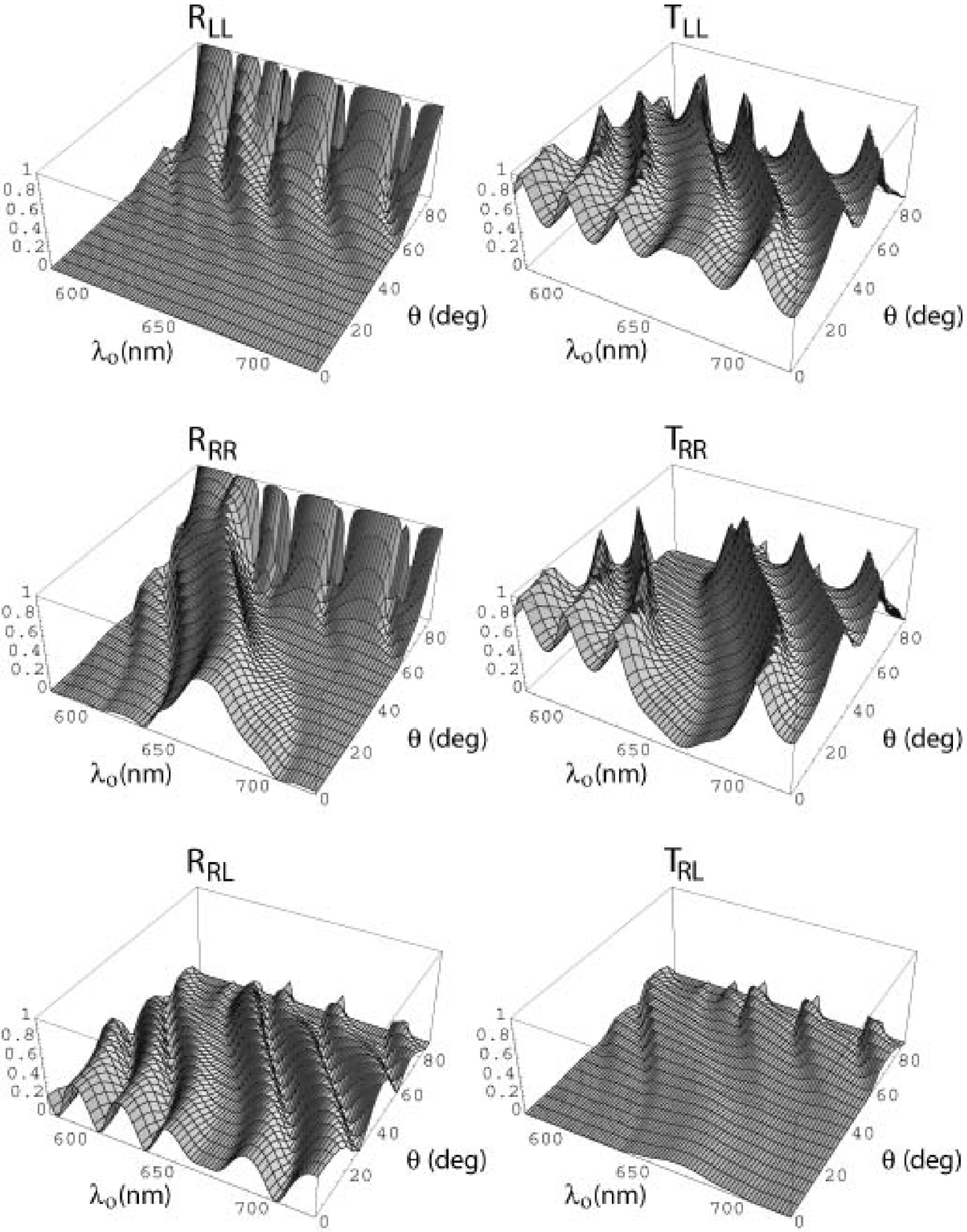,width=8cm }
\caption{Reflectances and transmittances of a locally biaxial SCM slab of thickness
$L=20\,\Omega$ as functions of the free--space wavelength $\lambda_o$
and the incidence angle $\theta$, when $\Ezdc=0$ and $\phi=0^\circ$. The local crystallographic
class 
of the SCM is orthorhombic $mm2$. Other parameters are: $\epsilon_1^{(0)} = 4.72$,$\epsilon_2^{(0)}= 5.20$,$\epsilon_3^{(0)}=5.43$, 
$r_{13}=34\times 10^{-12}$~m~V$^{-1}$,
$r_{23}=6\times 10^{-12}$~m~V$^{-1}$,
$r_{33}=63.4\times 10^{-12}$~m~V$^{-1}$,
$r_{42}=450\times 10^{-12}$~m~V$^{-1}$,
$r_{51}=120\times 10^{-12}$~m~V$^{-1}$,
all other $r_{JK}=0$,
$h=1$, $\Omega=150$~nm, and $\chi=90^\circ$. As   $T_{LR}=T_{RL}$   and $R_{LR}=R_{RL}$ to numerical accuracy, the plots of  $T_{LR}$ and
 $T_{LR}$ are not shown.}
\label{orthomm2-1}
\end{figure}

\begin{figure}[!h]
\centering \psfull
\epsfig{file=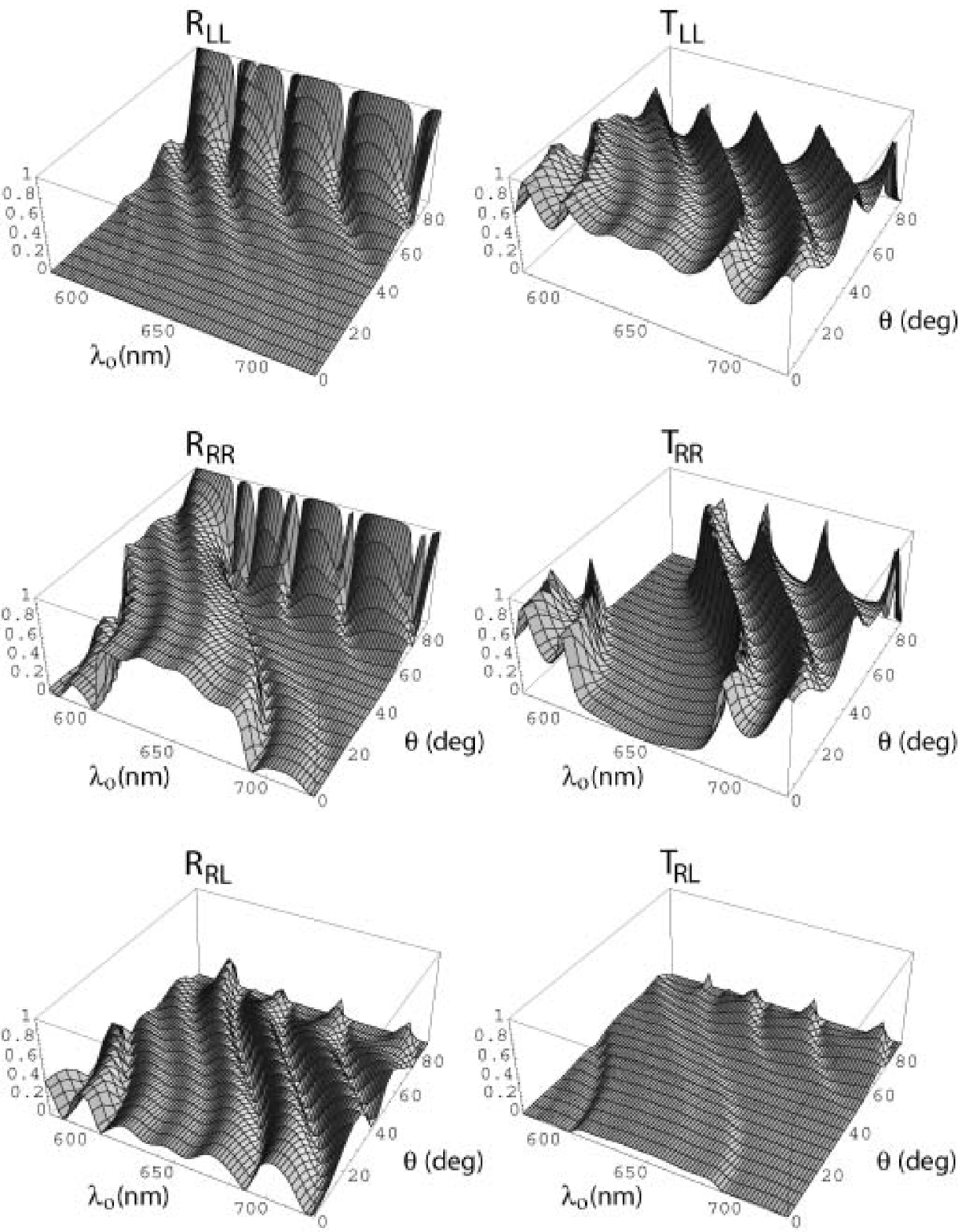,width=8cm }
\caption{Same as Fig. \ref{orthomm2-1}, except that a  dc voltage $V_{dc}= 2$~kV is applied
between the planes $z=0$ and $z=L$; thus, $\Ezdc= V_{dc}/L= 0.67$~GV~m$^{-1}$.
}
\label{orthomm2-2}
\end{figure}

Just like their locally uniaxial counterparts,
locally biaxial SCMs generally exhibit the CBP whether or not
$\Ezdc=0$. This is evident from Figs.~\ref{orthomm2-1} and \ref{orthomm2-2} 
which show the reflectance and transmittance
spectrums for  incidence angles $\theta\in\left[0^\circ,90^\circ\right)$
and $\phi=0^\circ$ when $\Ezdc=0$ and $\Ezdc=0.67$~GV~m$^{-1}$,
respectively. The chosen SCM has orthorhombic $mm2$
as its local crystallographic class, with the values of the relative permittivity scalars and the electro--optic coefficients the
same as for potassium niobate \cite{ZSB}. The plots of $R_{RR}$ 
and $T_{RR}$ in both figures show the Bragg regime
exhibiting a blueshift with decrease of $\cos\theta$. As the SCM slab is not very
thick ($L=20\Omega$), the CBP is not fully developed when $\Ezdc=0$,
but does exhibit the  broad top--hat profile of
the $R_{RR}$--ridge in Fig.~\ref{orthomm2-2} for  $\Ezdc\ne 0$. Clearly then, the application
of the dc voltage is efficacious in improving the CBP and confirms
the conclusion made in Section \ref{uni} that it would lead to thinner filters.

%%%%%%%%%%  Figure 7 begins %%%%%%%%%%%%
%%%%%%%%%%  Figure 7 ends  %%%%%%%%%%%%

%%%%%%%%%%  Figure 8 begins %%%%%%%%%%%%
%%%%%%%%%%  Figure 8 ends  %%%%%%%%%%%%

\begin{figure}[!h]
\centering \psfull
\epsfig{file=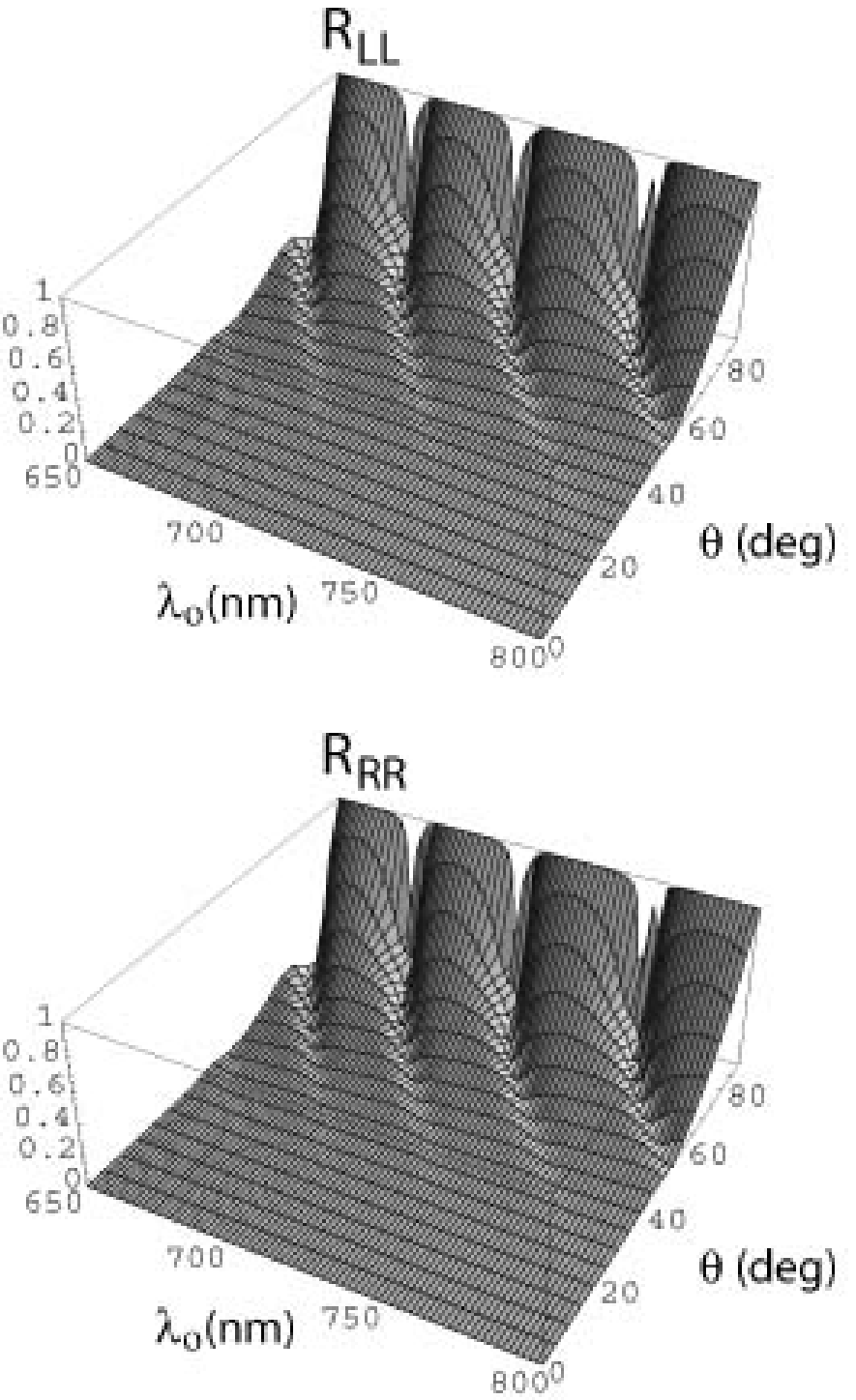,width=4cm }
\caption{Co--polarized reflectances  of a locally biaxial SCM slab when $V_{dc}= 0$.
All parameters are the same as for
Fig.~\ref{orthomm2-1}, except that $\chi=32.28^\circ$. Note the
absence of the CBP at normal and near--normal incidences at
the pseudoisotropic value chosen for $\chi$.
}
\label{orthomm2-3}
\end{figure}

\begin{figure}[!ht]
\centering \psfull
\epsfig{file=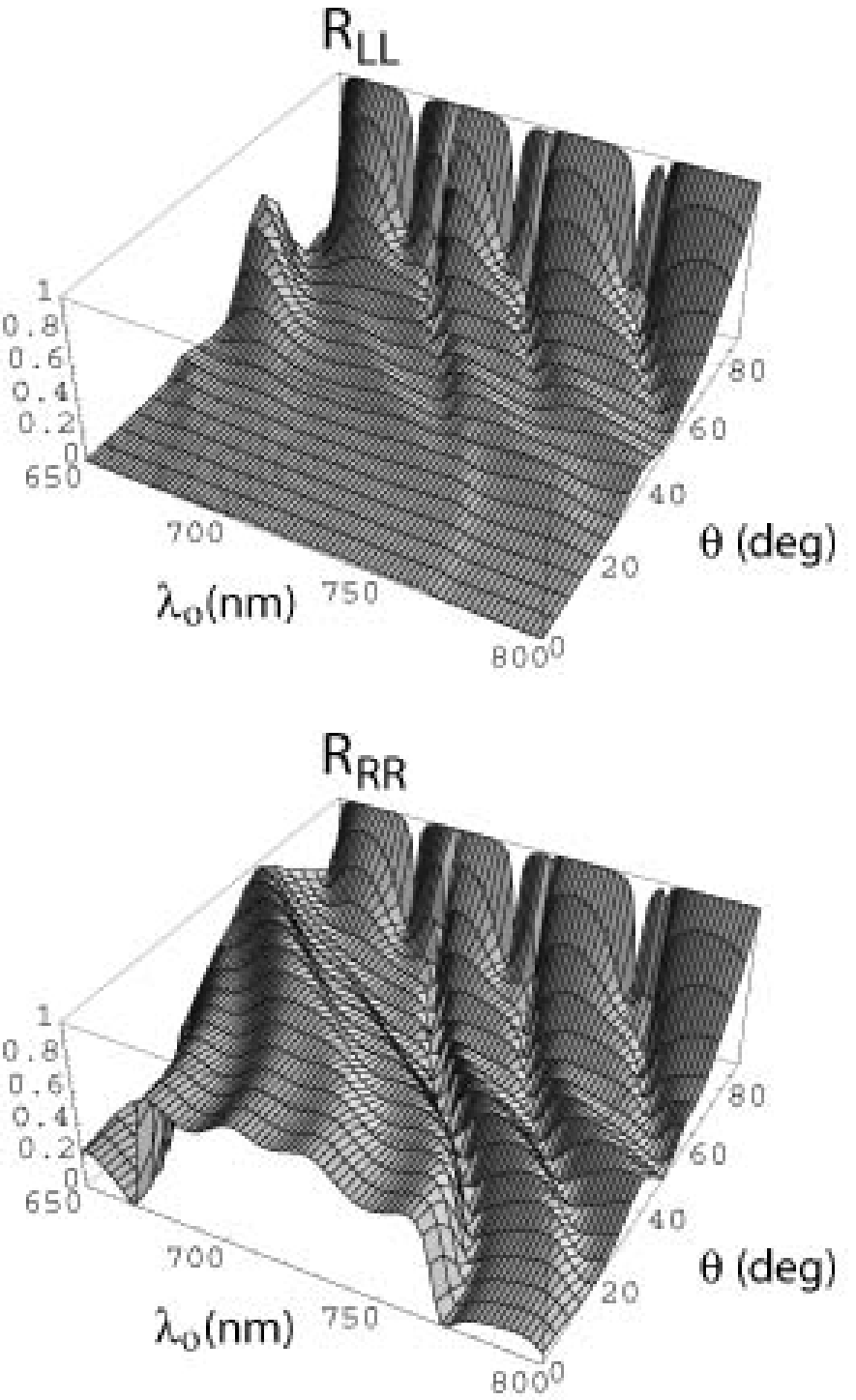,width=4cm }
\caption{Co--polarized reflectances  of a locally biaxial SCM slab when $V_{dc}= 2$~kV.
All parameters are the same as for
Fig.~\ref{orthomm2-3}. Note the
restoration of the CBP at normal and near--normal incidences at
the pseudoisotropic value chosen for $\chi$.
}
\label{orthomm2-4}
\end{figure}

When $\Ezdc=0$, locally biaxial SCMs can possess a
pseudoistropic point defined via \r{dd3}, whose influence is best seen for normal
incidence. For the relative permittivity scalars used to obtain the plots
of Fig.~\ref{orthomm2-1}, the pseudoisotropic point is identified by the
value $\chi=32.28^\circ$. Figures~\ref{orthomm2-3} 
and \ref{orthomm2-4} show spectrums
of the co--polarized reflectances $R_{RR}$ and $R_{LL}$ for the
same parameters as for Figs.~\ref{orthomm2-1}
and \ref{orthomm2-2}, respectively, except that $\chi=32.28^\circ$.
The Bragg regime is absent for $\theta$ less than at least $ 60^\circ$
when $\Ezdc=0$, but is restored in Fig.~\ref{orthomm2-4} by the
application of a dc voltage.

%%%%%%%%%%  Figure 9 begins %%%%%%%%%%%%
%%%%%%%%%%  Figure 9 ends  %%%%%%%%%%%%

%%%%%%%%%%  Figure 10 begins %%%%%%%%%%%%

%%%%%%%%%%  Figure 10 ends  %%%%%%%%%%%%

\subsection{Variation with $\phi$}

\begin{figure}[!ht]
\centering \psfull
\epsfig{file=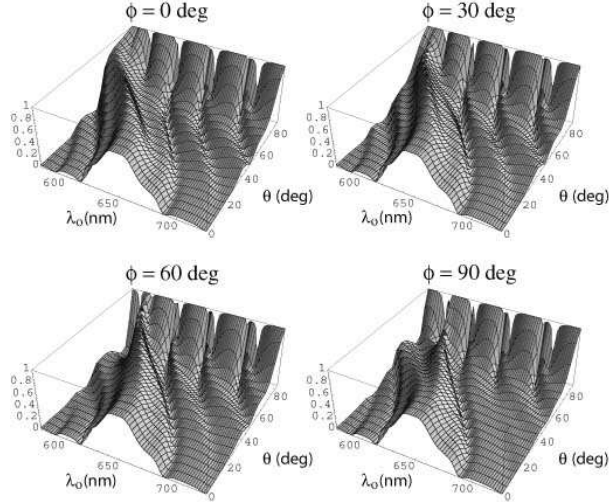,width=8cm }
\caption{Co--polarized reflectance $R_{RR}$ of a locally uniaxial SCM slab of thickness
$L=20\,\Omega$ as a function of the free--space wavelength $\lambda_o$
and the incidence angle $\theta$, when $\Ezdc=1.79$~GV~m$^{-1}$,
and $\phi=0^\circ$, $30^\circ$, $60^\circ$, and
$90^\circ$. The local crystallographic
class 
of the SCM is trigonal $3m$. Other parameters are: $\epsilon_1^{(0)} = \epsilon_2^{(0)}= 5.48$,$\epsilon_3^{(0)}=5.04$, 
$r_{22}=-r_{12}=-r_{61}=6.8
\times 10^{-12}$~m~V$^{-1}$, $r_{13}=r_{23}=9.6
\times 10^{-12}$~m~V$^{-1}$, $r_{33}=30.9\times 10^{-12}$~m~V$^{-1}$,
$r_{42}=r_{51}=32.6\times 10^{-12}$~m~V$^{-1}$, all other $r_{JK}=0$,
$h=1$, $\Omega=140$~nm, and $\chi=45^\circ$.  }
\label{trigonal3m-4}
\end{figure}

All plots shown heretofore are for $\phi=0^\circ$. In general, there is some effect of
$\phi$ on the reflectance and transmittance spectrums even in the absence
of the Pockels effect, which can be attributed to the relative orientation
of the incident electric field ${\bf e}_{inc}(0)$ with the principal components of
the projection of $\bar{\epsilon}^{SCM}(0)$ on the plane $z=0$ \c{VL98}.
Similar variations are found when $\Ezdc\ne 0$, but do not affect
the exhibition of the CBP~---~as may be gleaned from the plots
of $R_{RR}$ of a locally uniaxial SCM slab presented in Fig.~\ref{trigonal3m-4}.

%%%%%%%%%%  Figure 11 begins %%%%%%%%%%%%

%%%%%%%%%%  Figure 11 ends  %%%%%%%%%%%%

\subsection{Electrical manipulation of the CBP}
Several of the spectrums presented clearly show that the Bragg regime
not only blueshifts but also
narrows as $\cos\theta$ decreases, in line with previous reports
on non--electro--optic SCMs \cite{VLprsa}. Estimates of the
blueshift and narrowing can be obtained from curve--fitting
exercises \cite{VLepsa}
The case of normal incidence ($\theta=0^\circ$) therefore serves as
a bellwether for oblique incidence; the former is also significant in its own
right, because it represents possibly the most useful configuration
for optics. Therefore, as may be guessed
from Section \ref{axprop}, it is fortuitous that closed--form expressions
can be derived for important quantities for the normal--incidence
case.

In the present context of the manipulation of the CBP by the
application of a dc electric field, the two most important quantities are the
center--wavelength $\lambdao^{Br}$ and the FWHM bandwidth
$(\Delta\lambda_o)^{Br}$ of the Bragg regime for normal incidence,
as defined in \r{Br-def} and \r{dBr-def}. Let us note that
\begin{equation}  \label{Br0-def}
\lambdao^{Br}\Bigg\vert_{\Ezdc=0} = \Omega \left( \sqrt{\epsilon_{2}^{(0)}}+ \sqrt{\epsilon_{d}}%
\right)\,
\end{equation}
and
\begin{equation}  \label{dBr0-def}
(\Delta\lambda_o)^{Br}\Bigg\vert_{\Ezdc=0} =2\Omega 
\Big\vert\sqrt{\epsilon_{2}^{(0)}}- \sqrt{%
\epsilon_{d}}\Big\vert\,,
\end{equation}
which expressions clearly indicate that, if the Pockels effect is
not exploited, then
\begin{itemize}
\item[(i)] locally isotropic SCMs do not exhibit the CBP for all $\chi$,
\item[(ii)] locally uniaxial SCMs do not exhibit the CBP if $\chi=90^\circ$, and
\item[(iii)] locally biaxial SCMs do not exhibit the CBP if $\chi$
satisfies the pseudoisotropic condition \r{dd3}.
\end{itemize}
Against this backdrop, the role of the Pockels effect on the CBP
can be assessed analytically.

Table 1 shows the dependencies of $\epsilon_E$, $\epsilon_B$,
and $\epsilon_D$ on $\chi$, for all 20 local crystallographic classes.
In addition, $\epsilon_E$ may be a function
of $r_{41}$, $r_{43}$, $r_{61}$, and $r_{63}$; $\epsilon_B$
of $r_{21}$,   and $r_{23}$; and $\epsilon_D$ of $r_{11}$, $r_{13}$,
$r_{31}$, $r_{33}$, $r_{51}$, and $r_{53}$. Finally, $\epsilon_E$, $\epsilon_B$,
and $\epsilon_D$ vary linearly with $\Ezdc$. To second order in $\Ezdc$
then,
\begin{eqnarray}
\nonumber
&&
\sqrt{\epsilon_{B\xi}}\approx
\sqrt{\epsilon_2^{(0)}} + f_1(r_{21},r_{23},\chi)\,\Ezdc
\\
&&\qquad
+f_2(r_{21},r_{23},r_{41},r_{43},r_{61},r_{63},\chi)\,\left(\Ezdc\right)^2\,
\end{eqnarray}
and
\begin{eqnarray}
\nonumber
&&\sqrt{\epsilon_{D\xi}}\approx
\sqrt{\epsilon_d} + f_3(r_{11},r_{13},r_{31},r_{33},r_{51},r_{53},\chi)\,\Ezdc
\\
\nonumber
&&\qquad
+f_4(r_{11},r_{13},r_{31},r_{33},r_{41},r_{43},r_{51},r_{53},r_{61},r_{63},\chi)\\
&&\qquad\quad \times\,\left(\Ezdc\right)^2\,,
\end{eqnarray}
where $f_1$ to $f_4$ are functions of the identified electro--optic coefficients and
the tilt angle. Substitution of the foregoing expressions
in \r{Br-def} then leads to the following five statements:
\begin{itemize}
\item[A.] For both classes of locally isotropic SCMs, the shift of $\lambdao^{Br}$ 
does not depend on $\Ezdc$ but on 
$\left(\Ezdc\right)^2$.

\item[B.] For four classes of locally uniaxial SCMs, the shift of $\lambdao^{Br}$ 
does not depend on $\Ezdc$ but on 
$\left(\Ezdc\right)^2$. The four classes are tetragonal $422$, tetragonal ${\bar 4}2m$,
hexagonal $622$, and hexagonal ${\bar 6}m2$.

\item[C.] For the remaining nine classes of locally uniaxial SCMs, the shift of $\lambdao^{Br}$ 
depends on both $\Ezdc$ and
$\left(\Ezdc\right)^2$.

\item[D.] For three classes of locally biaxial SCMs, the shift of $\lambdao^{Br}$ 
does not depend on $\Ezdc$ but on 
$\left(\Ezdc\right)^2$. The four classes are orthorhombic $222$, monoclinic
$2$, and monoclinic $m$. 

\item[E.] For the remaining two classes of locally biaxial SCMs, the shift of $\lambdao^{Br}$ 
depends on both $\Ezdc$ and $\left(\Ezdc\right)^2$.  
\end{itemize}
Statements A to E for a {\it shift}
in the center--wavelength $\lambdao^{Br}$ upon the application of
a dc voltage also hold
true for the concurrent {\it change} in the FWHM bandwidth $(\Delta\lambda_o)^{Br}$.

Therefore, the center--wavelengths of the Bragg
regimes of SCMs of 11 local crystallographic classes will {\it shift},
and the FWHM bandwidths of the same will {\it change},
on the application of moderate dc voltages; whereas those of SCMs
of the remaining nine 
local crystallographic classes will require the application of higher
dc voltages to shift. Furthermore, the Bragg regimes will either redshift or blueshift, depending on the sign of $\Ezdc$, for the 11 local crystallographic classes;
but the shifts of the Bragg regimes will be insensitive to the sign of
$\Ezdc$, for the remaining nine local crystallographic classes.
These conclusions were verified by comparing the
Bragg--regime spectrums
for   trigonal $3m$, orthorhombic $mm2$, and tetragonal ${\bar 4}2m$ (not
shown here) classes.

The effect of the tilt angle $\chi$ on the FWHM bandwidth of
the Bragg regime for normal incidence is quite modified by the application of $\Ezdc$. Statements
(i) to (iii) are replaced as follows: When $\Ezdc\ne 0$,
\begin{itemize}
\item[I.] locally isotropic SCMs do not exhibit the CBP only if $\chi=45^\circ$;
\item[II.] locally uniaxial SCMs do not exhibit the CBP if $\chi=90^\circ$,
provided the local crystallographic class is neither tetragonal $\bar{4}$ nor
tetragonal $\bar{4}2m$; and
\item[III.] locally biaxial SCMs exhibit the CBP even if $\chi$ satisfies
the pseudoisotropic condition \r{dd3}.
\end{itemize} 
When $\chi=0^\circ$, the center--wavelength of the Bragg regime does
not shift for the following local crystallographic classes: tetragonal $4mm$,
hexagonal $6mm$, hexagonal ${\bar 6}m2$, trigonal $3m$, 
and orthogonal $mm2$. Likewise, when $\chi=90^\circ$, the center--wavelength of the Bragg regime does
not shift for the following local crystallographic classes:  tetragonal $422$,
hexagonal $622$, hexagonal $\bar{6}$, hexagonal ${\bar 6}m2$, and trigonal $32$.

\section{Concluding remarks}
A comprehensive treatment of the response
characteristics of a slab  of an electro--optic structurally chiral medium to normally as well
as obliquely incident plane waves was undertaken in this
paper. The SCM slab is endowed with one of 20 classes
of
point group symmetry, and is subjected to a dc voltage across its thickness. The
boundary--value problem was cast in the form of
a 4$\times$4 matrix ordinary
differential equation, whose solution yielded  the reflectances and transmittances of the SCM slab. The improvement~---~and, in some instances, the creation~---~of the circular Bragg phenomenon by the application of the dc voltage was theoretically demonstrated
and predicted to have either switching or circular--polarization--rejection
applications in optics. The possibility of thinner filters and electrical manipulation
of the CBP, depending on the local crystallographic class as well as 
the constitutive parameters of the SCM, were established. This comprehensive
study is expected to provide impetus to experimental research, possibly on ambichiral
versions of SCMs \cite{Lpla2006}.

Before concluding, let us contextualize the foregoing work in optics today.
SCMs can be considered to be one--dimensional photonic crystals (PCs). PCs have by now reached a mature stage of development, with their optical response characteristics well--understood and with many actual and potential applications \cite{PC1,PC2}. A recent trend concerns tunable or active PCs, whose optical response characteristics can change by means of some external agent. One way is to change their structural properties, for instance, by the application of mechanical stress \cite{YKOK,KG}. Another way is
to change their electromagnetic constitutive parameters, e.g., by incorporating the magnetically tunable ferroelectric and ferromagnetic materials \cite{FGV,ZSP,LDLSLR}
or by infiltrating a PC with the electrically controllable liquid crystals \cite{CLYG,HRR}.
We have shown here that the Pockels effect can be similarly exploited.

\vspace{0.2cm}

\noindent{\bf Acknowledgments.}
 We thank Partha P. Banerjee (University of Dayton,
OH, USA) and Venkat Gopalan (Pennsylvania State University, PA, USA)
for references on electro--optic materials.

\newpage

%%%%%%%%%%%%%%%%%%%%%%%%%%%%%%%%%%%%%%%%%

\noindent {\bf Table  1} Dependencies of $\epsilon_E$, $\epsilon_B$,
and $\epsilon_D$ on $\chi$, for the 20 classes of local
crystallographic symmetry. The coefficients $a_1$
to $a_{16}$ depend on various electro--optic coefficients.
 \\
\begin{small}

\begin{tabular}{|l||c|c|c|}
\hline
local crystallographic &  $\epsilon_E\,\epsilon_d$ & $\epsilon_B-\epsilon_2^{(0)}$
& $(\epsilon_D-\epsilon_d)\epsilon_d^2$
\\
class &    &  & 
\\ \hline\hline
cubic $23$ & $\cos(2\chi)$ & 0 & 0 
\\ 
cubic ${\bar 4}3m$ & $\cos(2\chi)$ & 0 & 0 
\\ \hline
tetragonal $4$ & $\cos^2\chi$ & $\sin\chi$ & $[a_{10}+a_{11}\cos(2\chi)]\sin\chi$
\\
tetragonal $4mm$ & 0&$\sin\chi$ &$[a_{10}+a_{11}\cos(2\chi)]\sin\chi$
\\
tetragonal $422$ & $\cos^2\chi$ & 0 & 0 
\\
tetragonal $\bar{4}$ & $a_1+a_2\cos(2\chi) $& $\sin\chi$ & $[a_{13}+a_{14}\cos(2\chi)]\sin\chi$
\\
tetragonal ${\bar 4}2m$ & $a_1+a_2\cos(2\chi) $& 0 & 0
\\
hexagonal $6$ & $\cos^2\chi$ & $\sin\chi$ & $[a_{10}+a_{11}\cos(2\chi)]\sin\chi$
\\
hexagonal $6mm$ & 0&$\sin\chi$ &$[a_{10}+a_{11}\cos(2\chi)]\sin\chi$
\\
hexagonal $622$ & $\cos^2\chi$ & 0 & 0 
\\
hexagonal ${\bar 6}$ &$\sin(2\chi)$& $\cos\chi$ & $\cos\chi\,\sin^2\chi$
\\
hexagonal ${\bar 6}m2$ &$\sin(2\chi)$& 0 & 0
\\
trigonal $3$ &  $a_4\cos\chi+a_5\sin\chi$& $a_6\cos\chi+a_{7}\sin\chi$&$[a_{{10}}+a_{{11}}\cos(2\chi)+a_{12}\sin(2\chi)]\sin\chi$
\\
trigonal $3m$ & $\sin(2\chi)$& $\sin\chi$ &$[a_{{10}}+a_{{11}}\cos(2\chi)]\sin\chi$
\\
trigonal $32$ & $\cos^2\chi$ & $\cos\chi$ & $\cos\chi\,\sin^2\chi$
\\ \hline
orthorhombic $222$ &  $a_1+a_2\cos(2\chi) $& 0 & 0
\\
orthorhombic $mm2$ & 0& $\sin\chi$ &$[a_{{10}}+a_{{11}}\cos(2\chi)]\sin\chi$
\\
monoclinic $2$ &  $a_1+a_2\cos(2\chi)+a_3\sin(2\chi)$& 0 & 0
\\
monoclinic $m$ & $a_1+a_2\cos(2\chi)+a_3\sin(2\chi)$& 0 & 0
\\
triclinic $1$ & $a_1+a_2\cos(2\chi)+a_3\sin(2\chi)$ &
$a_{8}\cos\chi+a_9\sin\chi$&
$[a_{{10}}+a_{{11}}\cos(2\chi)]\sin\chi$
\\
& & & $\quad+a_{15}\cos\chi + a_{16}\cos (3\chi)$\\
\hline

\end{tabular}
\end{small}

%%%%%%%%%%%%%%%%%%%%%%%%%%%%%%%%%%%%%%%%%

\end{document}